\documentclass{article}

\newif\ifarxiv
\arxivtrue          

\usepackage{silence}
\ifarxiv
  \usepackage[preprint]{neurips_2026}
\else
  \usepackage{neurips_2026}
\fi

\usepackage[utf8]{inputenc}
\usepackage[T1]{fontenc}

\usepackage{amsmath}
\usepackage{amssymb}
\usepackage{amsthm}
\usepackage{mathtools}

\usepackage[table]{xcolor}
\usepackage{graphicx}
\usepackage{svg}
\usepackage{subcaption}

\usepackage{booktabs}
\usepackage{longtable}
\usepackage{multirow}

\usepackage{algorithm}
\usepackage{algorithmic}

\usepackage{enumitem}
\usepackage{hyperref}
\usepackage{listings}
\usepackage{microtype}
\usepackage{xurl}

\newenvironment{widefigure}[1][]{\begin{figure}[#1]}{\end{figure}}
\newenvironment{widetable}[1][]{\begin{table}[#1]}{\end{table}}

\theoremstyle{plain}

\theoremstyle{definition}

\theoremstyle{remark}

\title{Caesar: Deep Agentic Web Exploration\\for Creative Answer Synthesis}
\author{%
  Jason Liang\thanks{Corresponding author: \texttt{jason.liang@cognizant.com}}, \quad
  Elliot Meyerson, \quad
  Risto Miikkulainen\thanks{Also affiliated with the University of Texas at Austin.} \\
  Cognizant AI Lab \\
}

\begin{document}

\maketitle

\begin{abstract}
To advance from passive retrieval to creative discovery of new ideas, autonomous agents must be capable of deep, associative synthesis. However, current agentic frameworks prioritize convergent search, often resulting in derivative summaries that lack creativity. Caesar is an agentic architecture designed to bridge the gap between information gathering and synthesis of new insights. Unlike existing agents that treat the web as a flat sequence of disconnected documents, Caesar performs a deep web traversal to construct a dynamic knowledge graph. This graph then serves as a navigational scaffold, guiding the agent to diverse, non-obvious information that flat retrieval would never encounter. Caesar thus consists of two components: (1) exploration driven by a dynamic context-aware policy that maximizes information coverage across the web's topological structure, and (2) synthesis through adversarial refinement that actively seeks novel perspectives rather than confirming established priors. Caesar demonstrates the ability to generate artifacts and answers characterized by high novelty and structural coherence, achieving 13--23\% improvement over state-of-the-art deep research agents in creative synthesis challenges, with strong dominance across all output formats.
\end{abstract}

\section{Introduction}
\label{sec:intro}

The advancement of technology has rarely been driven by the mere accumulation of facts, but rather by the novel synthesis of existing knowledge into new paradigms. From the combinatorial insights that birthed the steam engine to the cross-disciplinary reasoning underlying modern immunology \citep{nagy2013history, dickinson2022short}, human progress is defined by the ability to bridge disparate conceptual islands. To advance beyond rote instruction following, which recent frameworks already do well \citep{deng2023mind2web, zhou2024webarena, plaat2025agentic}, agents must evolve beyond the role of passive librarians. They must become active explorers capable of the deep, associative inquiry that characterizes true expertise.

Furthermore, despite the capabilities of Large Language Models \citep[LLMs;][]{brown2020language}, current web agents are still mostly performing stateless retrieval. Architectures built on standard retrieval-augmented generation \citep[RAG;][]{lewis2020rag} or linear ReAct loops \citep{yao2023react} effectively treat the internet as a flat sequence of disconnected documents. They optimize for precision by retrieving the most probable answer to a known question, but do not explore beyond it.

This limitation rules out open-ended inquiry, where the goal is not just summarization, but the generation of potentially useful information. This process leads to artifacts that are \emph{creative}; that is, they are \textit{New} (exhibiting novelty and rarity), \textit{Useful} (maintaining viability), and \textit{Surprising} (demonstrating a subversion of expected trajectories \citep{simonton2012creativity}).

This paper focuses on tasks that require such creativity. They are referred to as \textit{challenges}, each deliberately constructed around a distinct, well-documented LLM failure mode (Appendix~\ref{app:challenge_dataset}) along the \textit{New}, \textit{Useful}, and \textit{Surprising} axes. While recent advances in LLM capabilities (such as extended context windows, improved reasoning chains, and instruction following) provide the foundation for more sophisticated agent architectures, these capabilities alone are insufficient for creative synthesis. Without an explicit mechanism to expand beyond flat retrieval, even the best LLMs default to derivative summarization when used in standard agentic frameworks.

When a human researcher explores a new field, they build a mental map to identify non-obvious connections \citep{pirolli1999ift, boden2004creative}. Current deep research architectures such as \citet{gptr2024} and \citet{li2025webthinker} do not use such topological memory. By processing web pages in isolation, they exhibit ``navigational amnesia,'' loop through redundant content \citep{webresearcher2025}, and produce generic summaries that lack the novelty and lateral logic required for high-value creativity.

This paper presents an agentic architecture called \textbf{Caesar} that mimics the human approach (Figure~\ref{fig:caesar_final}). Caesar operates on the premise that the path taken to find information is a source of insight and constructs a dynamic knowledge graph during traversal. By systematically traversing the web's link structure, the agent discovers and indexes information that is structurally connected but often semantically orthogonal to the original challenge, seeding its knowledge base with the diverse material required for combinatorial creativity~\citep{boden2004creative}.

Caesar operates through a two-phase cognitive cycle. First, it maps the information topology, using a knowledge-guided policy to actively seek out conceptual bridges between disparate topics. Second, it synthesizes answers not by summarizing the graph, but by interrogating it. Caesar employs an adversarial refinement loop designed to escape the basin of attraction of generic LLM outputs. By actively seeking out contradictions, the system produces artifacts that adhere to the tripartite definition of creativity: delivering global novelty, functional viability, and the surprise of a lateral leap. This process is based on four technical innovations, each validated through ablation studies (Section~\ref{sec:ablation}):

\begin{widefigure}[t]
\centering
\includegraphics[width=\linewidth]{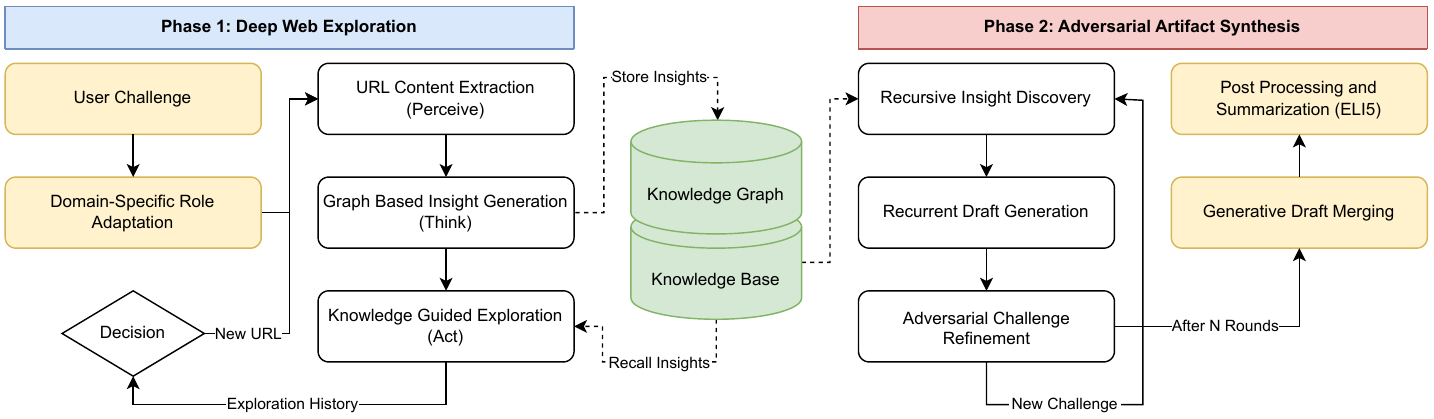}
\caption{\textbf{Visualization of the Caesar architecture.} \textbf{(Left) Phase 1: Deep Web Exploration.} A dynamic exploration policy controls a three-stage loop (Perceive, Think, Act) to traverse the web and to build a knowledge graph/database from insights. \textbf{(Right) Phase 2: Adversarial Artifact Synthesis.} Insights are retrieved to synthesize an initial draft. The agent then enters a recursive cycle, critiquing the current draft to generate adversarial challenges for refinement, before consolidating all versions via a generative merge and ELI5 summary. Together, these phases transform raw web traversal into a structured reasoning process that prioritizes creativity.}
\label{fig:caesar_final}
\end{widefigure}

(1) \textbf{Domain-Specific Role Adaptation (Section~\ref{sec:caesar_init}).} Caesar analyzes the challenge input to dynamically rewrite its own system prompt. This adaptation allows Caesar to adopt a persona specifically tuned to the domain's creative constraints, overcoming the safety-biased generic responses typical of models trained with Reinforcement Learning from Human Feedback \citep[RLHF;][]{ouyang2022training}.

(2) \textbf{Graph-Augmented Insight Generation (Section~\ref{sec:caesar_think}).} During traversal, Caesar constructs a knowledge graph that captures the topological structure of the information space. This graph serves dual purposes: it provides navigational context for the exploration policy (Section~\ref{sec:caesar_act}) and populates a semantic knowledge base with insights extracted along diverse traversal paths, ensuring the synthesis phase operates on a far richer information substrate than flat retrieval can provide.

(3) \textbf{Knowledge-Guided Exploration (Section~\ref{sec:caesar_act}).} Caesar introduces a dynamic decision-making policy for web traversal based on high-level meta-strategies. Unlike existing agents, Caesar utilizes exploration context and memory to detect whether navigation is stagnating, autonomously switching between depth-first expansion and strategic backtracking to maximize information gain while preventing endless looping or cycling.

(4) \textbf{Adversarial Artifact Synthesis (Section~\ref{sec:caesar_synthesis}).} This mechanism refines answers through recursive critique. Unlike static RAG, Caesar performs active gap analysis on intermediate drafts to formulate orthogonal challenges specifically targeting narrative weaknesses and contradictions. These divergent insights are then consolidated into a single cohesive artifact via a generative merge.

Caesar builds upon advances in autonomous web agents and deep research \citep{yao2023react, li2025webthinker, webresearcher2025}, graph-based memory and reasoning \citep{anokhin2024arigraph, adgr2025}, computational creativity \citep{boden2004creative, chakrabarty2024artifice}, iterative synthesis \citep{asai2023selfrag, shao2024storm}, and information foraging theory \citep{pirolli1999ift}. Sections~\ref{sec:caesar_explore} and~\ref{sec:caesar_synthesis} detail the technical mechanisms for exploration and synthesis, and Sections~\ref{sec:experiments} and~\ref{sec:discussion} present experimental results followed by discussion. An expanded review of related work is provided in Appendix~\ref{sec:related}.

\begin{figure*}[t]
\begin{minipage}[t]{0.5\linewidth}
\begin{algorithm}[H]
   \scriptsize
   \caption{Phase 1: Deep Web Exploration}
   \label{alg:caesar_exploration}
\begin{algorithmic}
   \STATE {\bfseries Input:} Challenge $Q$, Budget $T$
   \STATE {\bfseries Init:} $v_0 \leftarrow \textsc{Search}(Q)$, $\mathcal{S} \leftarrow [v_0]$, $G \leftarrow (V{=}\{v_0\}, E{=}\emptyset)$, $\mathit{KB} \leftarrow \emptyset$, $M \leftarrow \emptyset$, $L_f \leftarrow \emptyset$
   \STATE $\rho_r \leftarrow \textsc{GenerateRole}(Q, v_0)$
   \WHILE{$T > 0$ \AND $\mathcal{S} \neq \emptyset$}
       \STATE $v_c \leftarrow \mathcal{S}.\mathrm{peek}()$; $T \leftarrow T - 1$
       \STATE \COMMENT{1. Perceive (Sec.~\ref{sec:caesar_perceive})}
       \STATE $P_c, L \leftarrow \textsc{Extract}(v_c)$; $L_c \leftarrow L \setminus L_f$
       \STATE \textbf{if} $P_c$ is Invalid \textbf{then} $L_f \leftarrow L_f \cup \{v_c\}$; $\mathcal{S}.\mathrm{pop}()$; \textbf{continue}
       \STATE \COMMENT{2. Think (Sec.~\ref{sec:caesar_think})}
       \STATE $\mathcal{N}_c \leftarrow G.\mathrm{neighbors}(v_c)$
       \STATE $I_c \leftarrow \textsc{LLM}(P_c, \mathcal{N}_c, Q, \rho_r)$
       \STATE $G.\mathrm{update}(v_c, I_c)$; $\mathit{KB}.\mathrm{add}(I_c)$
       \STATE \COMMENT{3. Act (Sec.~\ref{sec:caesar_act})}
       \STATE $K_c \leftarrow \{\mathit{KB}.\mathrm{retrieve}(Q), M.\mathrm{recall}()\}$
       \STATE $a_m, v_n, Q', \text{trace} \leftarrow \pi_c(L_c, K_c)$
       \IF{$a_m = \textsc{Backtrack}$}
           \STATE $\mathcal{S}.\mathrm{pop}()$
       \ELSIF{$a_m = \textsc{WebSearch}$}
           \STATE $v_s \leftarrow \textsc{Search}(Q')$
           \STATE $\mathcal{S}.\mathrm{push}(v_s)$; $G.\mathrm{add}(v_c, v_s)$
       \ELSE[\emph{Explore}]
           \STATE $\mathcal{S}.\mathrm{push}(v_n)$; $G.\mathrm{add}(v_c, v_n)$
       \ENDIF
       \STATE $M.\mathrm{update}(v_c, a_m, v_n, \text{trace})$
   \ENDWHILE
   \STATE \textbf{Return} $G, \mathit{KB}$
\end{algorithmic}
\end{algorithm}
\end{minipage}%
\hfill
\begin{minipage}[t]{0.48\linewidth}
\begin{algorithm}[H]
   \small
   \caption{Phase 2: Adversarial Artifact Synthesis}
   \label{alg:caesar_synthesis}
\begin{algorithmic}
   \STATE {\bfseries Input:} Knowledge Base \emph{KB}, Rounds $N$, Challenge $Q_0$
   \STATE {\bfseries Init:} History $\mathcal{H} \leftarrow \emptyset$, $A_0 \leftarrow \emptyset$
   \FOR{$k=1$ {\bfseries to} $N$}
       \STATE \COMMENT{1. Recursive Insight Discovery}
       \STATE $I_k \leftarrow \textsc{GenerateInsightQA}(\mathit{KB}, Q_{k-1})$
       \STATE \COMMENT{2. Recurrent Draft Generation}
       \STATE $A_k, B_k \leftarrow \textsc{GenerateDraft}(I_k, A_{k-1})$
       \STATE $\mathcal{H} \leftarrow \mathcal{H} \cup \{(A_k, B_k)\}$
       \STATE \COMMENT{3. Adversarial Challenge Refinement}
       \STATE $Q_k \leftarrow \textsc{RefineChallenge}(A_k, Q_{k-1})$
   \ENDFOR
   \STATE \COMMENT{4. Generative Merge}
   \STATE $A_f \leftarrow \textsc{MergeDrafts}(\mathcal{H})$
   \STATE \COMMENT{5. Post-Processing (ELI5)}
   \STATE $A_e \leftarrow \textsc{PostProcess}(A_f)$
   \STATE \textbf{Return} $A_f, A_e$
\end{algorithmic}
\end{algorithm}
\end{minipage}
\end{figure*}

\vspace*{-0.5ex}
\section{Method for Phase 1 (Deep Web Exploration)}
\vspace*{-0.25ex}
\label{sec:caesar_explore}

As shown in Figure~\ref{fig:caesar_final}, Caesar is divided into two phases: (1) deep web exploration for knowledge collection and (2) exploitation of gained knowledge for answer synthesis. This section presents the core algorithms of the first phase.

\subsection{Deep Web Exploration Overview}
\vspace*{-0.25ex}
\label{sec:caesar_explore_overview}

Caesar treats exploration not as a linear sequence of retrieval steps, but as a stateful graph traversal problem. Formally, given a challenge $Q$, an exploration budget of $T$ steps, and a seed node $v_0$, the objective is to (1) construct a topological graph $G=(V,E)$, where nodes $v$ represent visited URLs and edges $e$ denote navigational transitions, and (2) populate a semantic knowledge base \emph{KB} with extracted insights. This dual-memory design explicitly decouples \textit{navigation} (managed by $G$) from \textit{information retention} (managed by \emph{KB}), allowing the agent to maximize information coverage before committing to a narrative structure.

The core of this phase is a recursive Perceive-Think-Act loop (Algorithm~\ref{alg:caesar_exploration}). Unlike traditional linear scrapers that operate on a stateless current-page basis, Caesar maintains a navigational stack $\mathcal{S}$ alongside the global graph $G$. This memory structure enables depth-first search capabilities: the agent can drill down into sub-topics until information gain plateaus, and then utilize the stack to backtrack and explore orthogonal branches. This mechanism mitigates the navigational amnesia characteristic of standard web exploration agents \citep{yao2023react, webresearcher2025}.

The process iterates until the budget $T$ is exhausted or the stack is empty. Upon termination, the graph $G$ is frozen, and the structured insights indexed in \emph{KB} are passed to the artifact synthesis module (Section~\ref{sec:caesar_synthesis}) for refinement.

\subsection{Domain-Specific Role Adaptation (Initialization)}
\label{sec:caesar_init}

Before the exploration loop (Algorithm~\ref{alg:caesar_exploration}), Caesar bootstraps the execution state by transforming the challenge $Q$ into a navigational entry point and a task-specific persona.

\textbf{Challenge Search Bootstrapping.} Rather than requiring a manual seed URL, Caesar generates its own starting node $v_0$. The agent expands $Q$ into auxiliary search terms, executes them via a web search API, and compiles the top search results into a synthetic HTML document. This document becomes the root node $v_0$ in graph $G$ and seeds the stack $\mathcal{S}$ with a starting point.

\textbf{Agent Role Generation.} The \textsc{GenerateRole} function adapts Caesar's role as defined in the LLM system prompt to match the task. By analyzing $Q$ and the initial search results in $v_0$, the agent synthesizes a specialized persona $\rho_r$ that defines an explicit goal and exploration philosophy. This approach ensures the reasoning in subsequent \textsc{Think} and \textsc{Act} phases aligns with the specific constraints of the challenge domain.

\subsection{URL Content Extraction (Perceive)}
\label{sec:caesar_perceive}

The \textsc{Perceive} function acts as the agent's sensory interface, accepting the current URL $v_c$ from the stack $\mathcal{S}$ and returning the extracted page content $P_c$ and candidate links $L_c$. To ensure robust retrieval against anti-bot countermeasures, the system replicates modern browser fingerprints and headers, preventing standard blocking on protected sources. The extraction method is content-agnostic, seamlessly parsing both HTML and PDF documents into plain text.

Similar to prior work \citep{deng2023mind2web}, the raw HTML or PDF is transformed into $P_c$ via a rigorous cleaning pass that only keeps the main text. Unnecessary elements like scripts and tags are removed to maximize the signal-to-noise ratio for processing. Simultaneously, the set of outgoing links $L_c$ is extracted. To ensure efficient traversal, $L_c$ is filtered to enforce user-specified domain boundaries while discarding URLs whose extraction failed ($L_f$), avoiding wasted budget on dead links. Beyond this structural pruning, the policy $\pi_c$ in the subsequent Act stage (Section~\ref{sec:caesar_act}) acts as a semantic discriminator: candidate links are ranked against the exploration context $K_c$, causing links that lack genuine semantic value (e.g., advertisements, navigation menus, SEO-optimized boilerplate) to be ranked lowest and effectively pruned from traversal.

\subsection{Graph-Augmented Insight Generation (Think)}
\label{sec:caesar_think}

To enrich the extracted insights beyond simple page summarization, Caesar conditions extraction on the local graph neighborhood. Insights $I_c$ are synthesized from the following stages:

\textbf{Topological Context Retrieval.} At node $v_c$, before analyzing the content $P_c$, the agent retrieves the semantic state of the local neighborhood $\mathcal{N}_c = \{v \in V : d_G(v, v_c) \leq k\}$. The experiments in this paper use $k=3$ by default and truncate to the nearest nodes, i.e., $|\mathcal{N}_c| \leq 30$. Following the paradigm of using graphs for agent memory \citep{anokhin2024arigraph}, this state includes insights associated with predecessor nodes and neighbors, creating a short-term memory window relative to $v_c$.

\textbf{Context-Aware Prompting.} The generation of $I_c$ is conditioned on the tuple $(P_c, \mathcal{N}_c, Q, \rho_r)$. The LLM is explicitly instructed to identify how the new content $P_c$ builds upon or challenges the retrieved context $\mathcal{N}_c$ rather than merely summarizing it. Similar to the approach in \citet{adgr2025}, this analysis enables Caesar to discover novel patterns and contradictions relative to its traversal path.

\textbf{Dual-State Storage.} The resulting insights $I_c$ are stored in a dual-memory system. First, $I_c$ is attached as an attribute to node $v_c$ in $G$, providing immediate context for future graph traversals. Second, $I_c$ is indexed in a vector store knowledge base \emph{KB}, enabling the agent to perform global semantic retrieval during the subsequent Act stage.

\subsection{Knowledge-Guided Exploration (Act)}
\label{sec:caesar_act}

To navigate in a manner that ensures continual progress, Caesar employs a dynamic policy $\pi_c$ that conditions action selection on a composite context $K_c$ derived from both the knowledge base \emph{KB} and episodic memory. It operates through the following steps:

\textbf{Dual-Context Retrieval.} The agent constructs $K_c$ by querying two functionally distinct sources. First, it searches \emph{KB}, a vector store of semantic insights extracted during the Think stage, for content relevant to $Q$. Second, it queries a persistent episodic memory store $M$, which records navigational actions (URLs visited, backtrack decisions, and reasoning traces) rather than page content. $M$ is queried using high-frequency keywords to surface historical navigation patterns, enabling $K_c$ to include past navigational failures or loops relevant to the current topic. Phase~2 (synthesis) operates exclusively over \emph{KB}; $M$ is used only during exploration.

\textbf{Meta-Strategy Formulation.} Based on the exploration history in $K_c$, the LLM selects a high-level meta-action $a_m \in \mathcal{A}$ using three mechanisms:
\begin{itemize}[leftmargin=*, nosep]
    \item \textsc{Explore}: Selects unvisited link $v_n \in L_c$ to expand the frontier, thereby deepening exploration of the current topic.
    \item \textsc{Backtrack}: Pops the stack $\mathcal{S}$ to return to the parent node $v_p$, escaping stagnant exploration regions.
    \item \textsc{WebSearch}: Retrieves new search results $v_s$ for $Q'$ (updated using $K_c$) to pivot exploration to a new area.
\end{itemize}

\textbf{Discriminative Link Selection.} If the meta-strategy is \textsc{Explore}, the agent selects the next link $v_n$ to visit from the candidate list $L_c$. The LLM acts as a discriminator, ranking $L_c$ based on $K_c$ and the meta-strategy. The agent then pushes $v_n$ to $\mathcal{S}$, adds the edge $(v_c, v_n)$ to $G$, and logs the move with reasoning trace into its memory $M$. After each action, $M$ is updated with a structured record $(v_c, a_m, v_n, \text{trace})$ enabling pattern-based detection of navigational loops. The action space is dynamically constrained at each step: \textsc{Backtrack} requires $|\mathcal{S}| > 1$, \textsc{WebSearch} is gated by remaining budget $S_m$, and \textsc{Explore} requires $|L_c| > 0$.

\section{Method for Phase 2 (Adversarial Artifact Synthesis)}
\label{sec:caesar_synthesis}

This section details the adversarial synthesis phase of Caesar and how it improves upon standard RAG, which typically relies on flat, single-shot retrieval. In contrast, Caesar is designed to emulate the recursive drafting and critiquing process of a human researcher. This process operates as a stateful recurrent system that performs active gap analysis to improve and fix a draft artifact over $N$ rounds of refinement. As outlined in Algorithm~\ref{alg:caesar_synthesis}, the adversarial artifact synthesis loop is composed of the following stages:

\textbf{Recursive Insight Discovery.}
Before synthesis begins, the agent executes a chain of inquiries to build a structured context window. Unlike standard RAG, which retrieves top documents in a single pass against the input challenge, Caesar employs a recursive probing strategy (\textsc{GenerateInsightQA}) over $\hat{T}$ iterations. Given a question $q_t$ (initialized as the challenge $Q_0$), Caesar uses \emph{KB} to retrieve relevant insights, generates an answer $a_t$, and then automatically creates a follow-up question $q_{t+1}$ to target the ambiguities (or gaps) in $a_t$. This process ensures that the context window contains a logical chain of reasoning rather than a bag of disjoint facts.

\textbf{Recurrent Draft Generation.}
In \textsc{GenerateDraft}, the generation of each draft artifact $A_k$ is conditioned on a composite context window $C_k = \{I_k, A_{k-1}\}$, where $I_k$ is the set of $(q, a)$ pairs generated above, and each answer $a$ is accompanied by source metadata retrieved directly from the vector knowledge base. In order to furnish the artifact with citations, Caesar creates a citation map $B_k$ that links every answer in $I_k$ to specific source URL indices. $C_k$ is used to prompt an LLM to generate $A_k$ by integrating the insights, previous draft, and citations.

\textbf{Adversarial Challenge Refinement.}
To prevent the agent from converging on a shallow summary, Caesar implements an active refinement loop via \textsc{RefineChallenge}. Between synthesis rounds, the system analyzes the artifact $A_k$ to identify narrative weaknesses. It then formulates a new, orthogonal challenge $Q_k$ explicitly prompted to target these weaknesses or contradictions in the current draft. $Q_k$ forces the agent to expand the exploration frontier in directions that maximize information gain relative to the current belief state.

\textbf{Generative Draft Merging.}
In this stage, \textsc{MergeDrafts} executes a generative unification of the complete draft history $\mathcal{H} = \{(A_k, B_k)\}_{k=1}^{N}$, consisting of the artifact $A_k$ and its citation map $B_k$ from each draft iteration. Rather than simply concatenating drafts, the system prompts an LLM to perform a high-level synthesis that selectively integrates the most relevant insights to create the final merged artifact $A_f$ that answers $Q_0$. The objective is to discover emergent patterns not visible in individual drafts and to construct a cohesive narrative that further develops the core strengths of previous drafts while actively addressing their weaknesses.

\textbf{Post-Processing and Summarization (ELI5).}
The output of a creative process that derives insights across diverse sources can be complex. To make the insights more understandable and appealing to a wider audience, the pipeline incorporates a \textsc{PostProcess} module that uses the ``Explain Like I'm 5'' (ELI5) paradigm \citep{fan2019eli5} to distill the final merged artifact into layperson-accessible language with optional token constraints. Crucially, this step is architecturally decoupled from the core synthesis loop, ensuring that the semantic simplification required for readability does not compromise the citation integrity or information density of the main artifact text.

\section{Experiments}
\label{sec:experiments}

To validate the efficacy of Caesar, it was compared against state-of-the-art research agents powered by the latest LLM models. Evaluations utilize a blinded LLM-as-a-Judge framework \citep{zheng2023judging} to assess performance on diverse challenges that exemplify the three dimensions of creativity: \textit{New}, \textit{Useful}, and \textit{Surprising}. The main results are summarized in Table~\ref{tab:main_results}.

\subsection{Solving Creative Challenges}
\label{sec:llm_judge_results}

\begin{widetable}[tb!]
\caption{\textbf{LLM-as-a-Judge Results.} Average scores for all agents under each output constraint. Cliff's Delta effect sizes~\citep{macbeth2011cliff} are uniformly large ($\delta \geq 0.76$, well above the $0.47$ large-effect threshold). $\delta=1.00$ corresponds to strict dominance, i.e., Caesar's lowest per-challenge mean exceeds the baseline's highest.}
\label{tab:main_results}
\centering\vspace{6pt}
\begin{small}
\begin{sc}
\begin{tabular}{lccccc}
\toprule
Agent & New & Useful & Surp. & Total & $\delta$ \\
\midrule
\multicolumn{6}{c}{\textbf{Full Answers (Unconstrained)}} \\
\addlinespace[2pt]
\textbf{Caesar} & \textbf{9.11} & \textbf{8.87} & \textbf{8.98} & \textbf{26.96} & -- \\
Gemini 3 (Deep) & 8.09 & 7.60 & 8.09 & 23.78 & 0.84 \\
Sonnet 4.5 (Deep) & 6.73 & 7.49 & 6.42 & 20.64 & 1.00 \\
GPT-5.2 (Deep) & 5.07 & 6.31 & 4.36 & 15.74 & 1.00 \\
Gemini 3 (Shallow) & 5.09 & 5.40 & 4.93 & 15.42 & 1.00 \\
Sonnet 4.5 (Shallow) & 5.27 & 4.40 & 5.04 & 14.71 & 1.00 \\
GPT-5.2 (Shallow) & 4.84 & 5.36 & 4.42 & 14.62 & 1.00 \\
\midrule
\multicolumn{6}{c}{\textbf{ELI5 Answers (Unconstrained)}} \\
\addlinespace[2pt]
\textbf{Caesar} & \textbf{8.62} & \textbf{8.69} & \textbf{8.38} & \textbf{25.69} & -- \\
Sonnet 4.5 (Deep) & 7.11 & 7.87 & 6.89 & 21.87 & 0.76 \\
Gemini 3 (Deep) & 5.89 & 5.78 & 6.02 & 17.69 & 1.00 \\
GPT-5.2 (Deep) & 5.38 & 6.71 & 5.00 & 17.09 & 1.00 \\
Sonnet 4.5 (Shallow) & 5.56 & 4.71 & 5.27 & 15.54 & 1.00 \\
GPT-5.2 (Shallow) & 4.82 & 5.11 & 4.31 & 14.24 & 1.00 \\
Gemini 3 (Shallow) & 3.96 & 4.47 & 3.58 & 12.01 & 1.00 \\
\midrule
\multicolumn{6}{c}{\textbf{ELI5 Answers (450 Word Limit)}} \\
\addlinespace[2pt]
\textbf{Caesar} & \textbf{8.33} & \textbf{8.69} & \textbf{8.11} & \textbf{25.13} & -- \\
Sonnet 4.5 (Deep) & 6.51 & 7.58 & 6.31 & 20.40 & 0.96 \\
Gemini 3 (Deep) & 6.64 & 6.20 & 6.82 & 19.66 & 0.92 \\
Sonnet 4.5 (Shallow) & 6.76 & 5.51 & 6.62 & 18.89 & 1.00 \\
GPT-5.2 (Shallow) & 4.89 & 5.51 & 4.40 & 14.80 & 1.00 \\
GPT-5.2 (Deep) & 4.40 & 6.07 & 3.71 & 14.18 & 1.00 \\
Gemini 3 (Shallow) & 4.27 & 5.09 & 3.91 & 13.27 & 1.00 \\
\bottomrule
\end{tabular}
\end{sc}
\end{small}
\end{widetable}

Caesar was evaluated against baseline agents using the following foundation models: Claude Sonnet 4.5 \citep{anthropic2025claude45}, GPT-5.2 \citep{openai2025gpt52}, and Gemini 3 Pro \citep{google2025gemini3}. Two configurations were tested: a \textit{shallow} agent utilizing standard single-step web search, and a \textit{deep} variant running a proprietary autonomous research mode that allows for unlimited web search steps (i.e.\ Research Mode for Claude, Deep Research for Gemini/GPT).

To ensure fairness, all agents (including Caesar) utilized the same basic prompt for generating answers with high reasoning effort. This prompt encourages insightful, interesting responses to challenges but is distinct from the judge rubrics to prevent overfitting and reward hacking. A panel of three LLM judges (Claude Sonnet 4.5, GPT-5.2, and Gemini 3 Pro) evaluated anonymized answers on a 10-point scale across three dimensions: \textit{New} (Novelty/Rarity), \textit{Useful} (Viability/Alignment), and \textit{Surprising} (Non-obvious connections). Agent prompts and judge rubrics are described in Appendices~\ref{app:core_prompts} and~\ref{app:experiment_setup}, while implementation details and hyperparameters are listed in Appendix~\ref{app:implementation}.

Each agent was scored across five challenges designed to test different aspects of creativity: Constrained Synthesis, Counterfactual Reasoning, Cross-Domain Synthesis, Meta-Creativity, and Open-Ended Synthesis. These aspects were identified by searching prior work for common creativity failure modes of LLMs (Appendix~\ref{app:challenge_dataset}). Each challenge was evaluated under three output constraints: (1) unconstrained full answers, (2) unconstrained ELI5 summaries, and (3) length-constrained ELI5 summaries (450 words). Every judge scored each challenge three times to reduce evaluation noise. To better understand how Caesar outperforms the baselines, analysis of sample answers and detailed challenge scores are provided in Appendices~\ref{app:qualitative_comparison} and~\ref{app:detailed_results} respectively.

As shown in Table~\ref{tab:main_results}, Caesar significantly outperformed all baselines across all experimental settings, achieving strictly large Cliff's Delta effect sizes ($\delta > 0.47$) against all baselines across all output formats \citep{macbeth2011cliff, meissel2024using}. For these calculations, each of the challenges was treated as a single, independent data point. In the unconstrained full answer setting, Caesar achieved a total score of 26.96, surpassing the runner-up (Gemini 3 Deep Research) by a margin of 3.18 points. Notably, Caesar demonstrated the highest scores in the \textit{New} (9.11) and \textit{Surprising} (8.98) metrics. As further elaborated in Appendix~\ref{sec:graph_topology}, this result suggests that graph-guided exploration fosters greater creativity than standard search and retrieval, which often defaults to derivative summarization.

The performance gap persisted with semantic compression. In the unconstrained ELI5 setting, where significantly simpler answers were required, Caesar achieved a total score of 25.69 compared to the runner-up's 21.87. While baseline models typically sacrificed surprise to meet the readability constraints of the ELI5 persona, Caesar retained a high surprise score of 8.38, indicating that the simplification process did not dilute the novelty of the retrieved insights.

This trend persisted even in the length-constrained ELI5 (450 words) task. Caesar maintained a high novelty score of 8.33 while peer models degraded in performance. As discussed in Appendix~\ref{sec:iter_convergence}, this result validates the architectural decision to perform draft refinement followed by a generative merge. Caesar maximizes semantic density, ensuring that high-utility insights are retained even when narrative flair is pruned. This superior length-constrained performance shows that Caesar's creative capabilities do not follow simply from generating longer answers to challenges, but instead from the actual content in the answers.

\textbf{Compute-Controlled Comparison.} To control for computational expenditure, Caesar was run at $T{=}250$ using GPT-5-mini, matching the estimated \$5 per-challenge cost of Gemini Deep Research (Appendix~\ref{app:cost_controlled}). Under this budget, Caesar (26.16) still significantly outperformed all deep research baselines (Gemini 3: 24.37, Sonnet 4.5: 21.00, GPT-5.2: 16.16) with large effect sizes ($\delta \geq 0.76$), demonstrating that Caesar's advantage is not merely a function of increased compute.

\textbf{Human Evaluation Study.} To benchmark the automated evaluation, 23 human raters compared length-controlled, anonymized summaries from Caesar and the strongest baseline (Gemini 3 Deep Research) in pairwise A/B matchups. Raters scored under the same NUS rubric used by the LLM judges. Human raters preferred Caesar in 63 of 112 comparisons (56.25\%, odds ratio 1.29), supporting the LLM-judge findings (Appendix~\ref{app:human_eval}).

\subsection{Judge Bias Analysis}
\label{sec:judge_bias}

\begin{widetable}[tb!]
\caption{\textbf{Judge Bias and Robustness Analysis.} \textit{Left:} Self-preference biases across all judges. Positive values indicate a preference for models of the same family. \textit{Right:} Average scores with the GPT-5.2 judge excluded (Sonnet and Gemini judges only), addressing potential circularity since Caesar uses GPT-5.2 as its foundation model. Caesar is the best-performing agent.}
\label{tab:bias_results}
\centering\vspace{6pt}
\begin{sc}
\resizebox{\linewidth}{!}{%
\begin{tabular}{lccc}
\toprule
\textbf{Judge} & \textbf{Full} & \textbf{ELI5} & \textbf{450w} \\
\midrule
Gemini & +1.35 & -0.12 & -0.29 \\
Claude & +0.98 & -0.20 & -0.22 \\
GPT    & -0.82 & -0.75 & +0.97 \\
\bottomrule
\vspace*{7.25ex}
\end{tabular}%
\hspace{2em}%
\begin{tabular}{lccc}
\toprule
\textbf{Agent (no GPT judge)} & \textbf{Full} & \textbf{ELI5} & \textbf{450w} \\
\midrule
\textbf{Caesar} & \textbf{27.27} & \textbf{25.94} & \textbf{25.10} \\
Gemini 3 (Deep) & 25.53 & 18.10 & 20.27 \\
Sonnet 4.5 (Deep) & 21.73 & 22.60 & 21.03 \\
GPT-5.2 (Deep) & 16.20 & 17.74 & 14.07 \\
Gemini 3 (Shallow) & 15.87 & 11.64 & 13.29 \\
Sonnet 4.5 (Shallow) & 15.27 & 16.00 & 19.37 \\
GPT-5.2 (Shallow) & 14.70 & 14.10 & 14.27 \\
\bottomrule
\end{tabular}%
}
\end{sc}
\end{widetable}

To ensure robust scoring, self-preference bias was analyzed (Table~\ref{tab:bias_results}, left) by comparing each judge's scores for its own model family against the other judges' scores. The Gemini judge exhibited a positive bias ($+1.35$) on full answers, the largest in magnitude across all judge-format pairs and consistent with LLM self-preference~\citep{panickssery2024llm}; the Claude judge similarly showed a positive bias on full answers ($+0.98$). To address potential circularity due to Caesar using GPT-5.2, Table~\ref{tab:bias_results} (right) reports results with the GPT-5.2 judge excluded (Caesar remains the top agent).

A separate verbosity analysis reveals a moderate positive correlation ($r = 0.47$) between output length and judge score across all agents. However, this is an instance of Simpson's paradox driven by tier differences: deep research agents naturally produce both longer and higher-quality outputs than shallow agents. Within the deep-research tier (Caesar's direct peers), the correlation is weakly negative ($r = -0.13$), confirming that Caesar's advantage stems from content quality, not length. Caesar's strong performance under the 450-word constraint (Table~\ref{tab:main_results}) further corroborates this finding.

\subsection{Ablation Studies}
\label{sec:ablation}

To isolate the contributions of each architectural component, multiple ablation studies were conducted. Results for the full answers are described in this section; results for ELI5 output constraints follow the same trends and are provided in Appendix~\ref{app:detailed_results}.

\begin{widetable}[tb!]
\caption{\textbf{Graph \& Single-Hop Ablation (Full Answers).} \textit{Left:} Removing the entire knowledge graph (replacing graph-guided exploration with flat top-$T$ web search results) produces a large negative effect. \textit{Right:} Reducing $\mathcal{N}_c$ to single-hop ($k{=}1$, $|\mathcal{N}_c|{=}5$) causes a similar effect. Scores are totals (\emph{New}+\emph{Useful}+\emph{Surprising}) for the final merged artifact $A_f$.}
\label{tab:ablation_graph}
\centering\vspace{6pt}
\begin{small}
\begin{sc}
\begin{tabular}{lcc}
\toprule
Variant & Total & $\delta$ \\
\midrule
Caesar (Control) & \textbf{25.11} & -- \\
No KG (Flat Search) & 23.76 & 0.52 \\
\bottomrule
\end{tabular}
\hspace{1em}
\begin{tabular}{lcc}
\toprule
Variant & Total & $\delta$ \\
\midrule
Caesar (Control) & \textbf{25.45} & -- \\
Single-Hop ($k{=}1$) & 23.69 & 0.52 \\
\bottomrule
\end{tabular}
\end{sc}
\end{small}
\end{widetable}

First, removing the entire knowledge graph and replacing it with flat top-$T$ web search results caused a substantial degradation ($\delta = 0.52$, large), confirming that the graph's value lies in \textit{guiding diverse exploration} rather than providing local context. Second, reducing $\mathcal{N}_c$ to single-hop neighborhood size ($k{=}1$, $|\mathcal{N}_c|{=}5$) similarly results in decreased performance ($\delta = 0.52$, large), suggesting that broader topological context is important for generating creative answers. These results are summarized in
Table~\ref{tab:ablation_graph}.

\begin{widetable}[tb!]
\caption{\textbf{Exploration Ablation and Draft Ablation (Full Answers)}. \textit{Left:} Reducing the exploration budget ($T$) substantially degrades all three creativity metrics. \textit{Right:} Adversarial refinement ($A_1 \rightarrow A_3$) maximizes \textit{Surprising} at the cost of \textit{Useful}; the generative merge ($A_f$) reconciles both.}
\label{tab:ablation_iterations}
\centering\vspace{6pt}
\begin{scriptsize}
\begin{sc}
\begin{tabular}{lcccc c}
\toprule
Config & New & Usef. & Surp. & Total & $\delta$ \\
\midrule
\textbf{1000 Iter} & \textbf{8.22} & \textbf{8.54} & \textbf{8.16} & \textbf{24.92} & -- \\
500 Iter & 7.46 & 7.80 & 7.48 & 22.74 & 0.76 \\
250 Iter & 7.22 & 7.48 & 7.14 & 21.84 & 0.92 \\
\bottomrule
\end{tabular}
\hspace{1em}
\begin{tabular}{lcccc c}
\toprule
Variant & New & Usef. & Surp. & Total & $\delta$ \\
\midrule
\textbf{Final} & \textbf{7.86} & \textbf{8.28} & 7.68 & \textbf{23.82} & -- \\
Draft 3 & 7.90 & 7.04 & \textbf{7.98} & 22.92 & 0.32 \\
Draft 1 & 6.58 & 7.62 & 6.22 & 20.42 & 0.92 \\
\bottomrule
\end{tabular}
\end{sc}
\end{scriptsize}
\end{widetable}

Third, reducing the exploration iterations from 1000 to 250 caused a substantial decline of $-3.08$ in total score ($\delta = 0.92$, large), supporting the hypothesis that non-obvious insights require deep topological traversal. Fourth, adversarial refinement ($A_3$) boosts \textit{Surprising} (+1.76) and \textit{New} but reduces \textit{Useful}. The generative merge ($A_f$) recovers utility (from 7.04 to 8.28) while retaining discovered insights, confirming that creativity benefits from divergent and convergent thinking~\citep{koivisto2023divergent}. These results are summarized in Table~\ref{tab:ablation_iterations}.

Overall, these ablation studies validate three design principles: (1) the knowledge graph's value lies in guiding diverse exploration with broader topological context, (2) deeper exploration discovers rarer insights, and (3) adversarial refinement escapes surface-level summarization. Additional ablations are provided in the Appendix~\ref{app:detailed_ablation}: per-draft breakdowns of the knowledge graph and single-hop ablations (Tables~\ref{tab:graph_ablation_full},~\ref{tab:multihop_ablation}), and exploration and draft ablations across all output constraints (Table~\ref{tab:ablation_iterations_eli5}).

\section{Discussion}
\label{sec:discussion}

Two architectural mechanisms underlie Caesar's performance, each documented qualitatively in the appendices. Graph-guided exploration produces topologies that vary with the challenge, ranging from breadth-first starbursts to depth-first chains (Appendix~\ref{app:visualization}). Adversarial synthesis then escapes generic summarization through recursive insight discovery, traced via a four-iteration case study that evolves an abstract concept into a verified operational model (Appendix~\ref{app:recursive_discovery}). Appendix~\ref{app:broader_impacts} discusses broader societal impacts and responsible-use considerations.

\subsection{Limitations}
\label{sec:limitations}

Caesar's performance gains (Table~\ref{tab:main_results}) require significantly more token usage and wall-clock time than a standard single-turn retrieval agent, aligning with recent observations that scaling test-time compute can yield intelligence gains analogous to scaling model parameters \citep{brown2020language}. Caesar's heavy token usage makes it less suitable for low-latency applications; it is optimized for deep research tasks where the value of a breakthrough insight justifies the computational expense. A full breakdown of computational costs is provided in Appendix~\ref{app:cost}.

Furthermore, despite its success, Caesar's architecture exhibits specific failure modes. First is the SEO trap: if the bootstrapping phase (Section~\ref{sec:caesar_init}) seeds the graph $G$ within a cluster of low-quality, search-optimized content, the graph-based reasoning can become trapped in a recursive loop of low-value information. Second is unnecessary complexity: when an input challenge requires only basic fact retrieval, the adversarial loop adds unjustifiable overhead. Third, on speculative tasks where the answer relies on parametric invention rather than retrievable evidence, the graph's overhead is not offset by retrieval benefits: on Open-Ended Synthesis, Caesar's lead is smallest in Full Answers, narrowly trails Sonnet 4.5 Deep in unconstrained ELI5, and effectively ties under the 450-word constraint (Tables~\ref{tab:detailed_full_answers},~\ref{tab:detailed_eli5_answers},~\ref{tab:detailed_eli5_450_answers}).

Additionally, the evaluation is limited to five carefully designed challenges targeting specific creativity failure modes (Appendix~\ref{app:challenge_dataset}). While the large Cliff's Delta effect sizes and consistent trends across all output formats support the reliability of these findings, a larger benchmark spanning more challenges and additional creative dimensions is needed to establish broader generalizability.

Finally, despite the multi-judge panel and corroborating human evaluation, LLM-as-a-Judge remains an imperfect proxy for human creativity assessment. Judges trained via similar procedures may share systematic blind spots, and the NUS rubric, while grounded in creativity theory~\citep{simonton2012creativity}, may not capture all dimensions of creative value. Additionally, the deep research baselines rely on proprietary autonomous modes whose internal behavior, token budgets, and retrieval strategies are not publicly documented, limiting exact reproducibility of the comparison. The compute-controlled experiment (Section~\ref{sec:experiments}) partially mitigates this confound.

\subsection{Future Work}

Future work will extend Caesar from a single agent to a collaborative multi-agent swarm, allowing agents to specialize in complementary exploration patterns before merging discoveries. The architecture also adapts naturally to domains beyond question answering, such as program synthesis on the ARC-AGI benchmark~\citep{chollet2019measure}. This new domain would test whether Caesar can generate useful solutions for hard reasoning problems. Finally, adaptive mechanisms that dynamically calibrate exploration depth based on challenge complexity would reduce the unnecessary overhead Caesar incurs on simpler challenges.

\section{Conclusion}
\label{sec:conclusion}

This paper presents Caesar, a framework that advances autonomous web search by combining graph-guided deep exploration with adversarial artifact synthesis. Ablation analysis reveals that the knowledge graph's primary value lies in guiding the agent to diverse, non-obvious information through macro-level navigation, while the adversarial refinement loop escapes the basin of generic summarization. Together, these mechanisms enable Caesar to bridge the gap between static retrieval and active discovery, significantly outperforming state-of-the-art research agents.

\pagebreak

\bibliographystyle{plainnat}
\bibliography{caesar}

\newpage
\appendix

{\bf\large APPENDICES}

The appendices expand the discussion on the main paper as follows:
\begin{itemize}[leftmargin=*] \itemsep0em
    \item \textbf{Appendix~\ref{sec:related}} provides a deep, in-depth review of related work.
    \item \textbf{Appendix~\ref{app:experiment_setup}} describes the experimental design, such as challenges, prompts, and judge rubrics.
    \item \textbf{Appendix~\ref{app:detailed_results}} presents detailed tabular results for all challenge categories and output constraints.
    \item \textbf{Appendix~\ref{app:qualitative_comparison}} contrasts Caesar with baselines through deep comparison of agent answers.
    \item \textbf{Appendix~\ref{app:visualization}} analyzes the importance of graph exploration and visualizes the knowledge graph topologies constructed during Phase 1.
    \item \textbf{Appendix~\ref{app:recursive_discovery}} analyzes the dynamics of adversarial synthesis and presents a case study of the recursive insight chains generated during Phase 2.
    \item \textbf{Appendix~\ref{app:detailed_ablation}} provides detailed ablation results to better understand the impact of each component.
    \item \textbf{Appendix~\ref{app:cost_controlled}} details the compute-controlled comparison experiment.
    \item \textbf{Appendix~\ref{app:human_eval}} explains the human evaluation study design and results.
    \item \textbf{Appendix~\ref{app:implementation}} outlines the system implementation details and hyperparameters.
    \item \textbf{Appendix~\ref{app:core_prompts}} lists the agent prompts for both the exploration and synthesis phases.
    \item \textbf{Appendix~\ref{app:cost}} analyzes the computational costs for a full experimental run.
    \item \textbf{Appendix~\ref{app:broader_impacts}} discusses broader impacts, dual-use risks, and responsible deployment considerations.
\end{itemize}


\section{Related Work}
\label{sec:related}

The evolution of autonomous web agents has shifted from rigid task execution to dynamic exploration \citep{yao2023react}, yet the ability to synthesize novel insights over long horizons remains elusive \citep{webresearcher2025}. This section reviews the progression from linear retrieval frameworks to emerging graph-based cognitive architectures, highlighting the specific limitations in navigational reasoning and insight consolidation that Caesar addresses.

\textbf{Autonomous Web Agents and Deep Research.}
The development of generalist web agents has been accelerated by realistic benchmarks such as WebArena \citep{zhou2024webarena} and Mind2Web \citep{deng2023mind2web}. While the ReAct framework \citep{yao2023react} successfully established a standard for interleaving reasoning and acting, it fundamentally relies on a linear interaction history, resulting in navigational loops and limited context in complex environments. To mitigate this problem, deep research architectures like WebThinker \citep{li2025webthinker} and GPT-Researcher \citep{gptr2024} introduced hierarchical planning. Most recently, WebResearcher \citep{webresearcher2025} advanced this paradigm via IterResearch, allowing agents to operate without fixed context windows. However, these systems decouple exploration from synthesis only in terms of memory management, not reasoning. Caesar extends this philosophy by decoupling the graph construction phase from artifact synthesis, ensuring that the agent maps the information topology fully before attempting to construct a narrative.

\textbf{Graph-Based Memory and Reasoning.}
To overcome the limitations of linear history, researchers have increasingly adopted graph-based memory structures. AriGraph \citep{anokhin2024arigraph} demonstrated that knowledge graphs significantly outperform vector-only approaches for episodic memory in complex environments. In 2025, the focus shifted from static retrieval to active reasoning. Agentic Deep Graph Reasoning \citep{adgr2025} showed how agents can actively generate concepts and merge them into a self-organizing global graph, while G-Memory \citep{gmemory2025} proposed hierarchical insight graphs for multi-agent coordination. Caesar adapts these concepts specifically for navigational state tracking. Unlike general-purpose graph memories, Caesar combines knowledge graph insights with navigation history to detect stagnation and to force backtracking, transforming the graph from a passive storage unit into an active control signal for exploration.

\textbf{Computational Creativity and Associative Search.}
A critical open challenge in LLM reasoning is to extend agents beyond summarization to creativity. While standard LLMs can produce novel outputs via high-temperature sampling or prompting strategies like Tree of Thoughts \citep{long2023tot}, these approaches often rely on stochastic randomization rather than structured reasoning. Critics argue that without architectural constraints, LLMs default to mean-seeking behavior, producing statistically probable but uninspired content \citep{chakrabarty2024artifice}. Caesar departs from these stochastic methods by implementing System 2 creativity \citep{kahneman2011thinking}: it replaces random sampling with structured chain-of-thought reasoning. By explicitly seeking gaps and bridges between distant insights, Caesar aligns with theories defining creativity as the combinatorial discovery of non-obvious semantic connections \citep{boden2004creative}, rather than mere random divergence.

\textbf{Iterative Synthesis and Refinement.}
Standard Retrieval-Augmented Generation \citep[RAG;][]{lewis2020rag} typically follows a single-pass paradigm that results in simple and shallow generated reports. Recent work has shifted towards iterative refinement to improve depth. Self-RAG \citep{asai2023selfrag} introduced reflection tokens to critique generations, while STORM \citep{shao2024storm} synthesized Wikipedia-like articles through multi-perspective questioning. However, STORM relied on pre-generated outlines, restricting the agent to filling known information slots. Caesar aligns more closely with the evolving strategies of Iterative RAG \citep{choi2025mirag} but introduces a dynamic challenge evolution mechanism. Rather than filling a static outline, Caesar's synthesis of one draft actively generates adversarial challenges for the next, allowing the narrative structure to emerge organically from the discovered data.

\textbf{Active Information Foraging.}
Theoretical foundations from Information Foraging Theory \citep{pirolli1999ift} are increasingly applied to model the trade-off between information gain and navigational cost \citep{qian2025inforage}. DeepResearch Eco \citep{souza2025deepresearch} implemented this approach via depth-controlled recursive exploration. Caesar is similar but enforces it via a dynamic policy that conditions actions on both the global knowledge graph and local history. This approach allows the agent to make economic decisions about when to abandon low-value navigational paths more effectively than purely heuristic approaches.

\section{Experimental Setup}
\label{app:experiment_setup}

This section details the experimental design, including the specific challenges selected to test distinct types of creativity. This section also provides prompts given to agents for answering challenges and the evaluation rubric utilized by LLM judges to assess performance across the Novelty, Usefulness, and Surprise metrics.

\subsection{Challenge Dataset}
\label{app:challenge_dataset}

The evaluation dataset consists of five different challenges, each chosen to represent a unique category for creativity. These categories were designed to test specific dimensions of creativity where standard LLMs notoriously struggle. The challenge designs draw directly from \citeauthor{guilford1967nature}'s \citeyearpar{guilford1967nature} distinction between convergent and divergent thinking, as well as \citeauthor{koestler1964act}'s \citeyearpar{koestler1964act} concept of bisociation, the creative act of connecting previously unrelated matrices of thought. The five challenges and their categorical structure were finalized prior to building Caesar's evaluation pipeline; no pilot results on Caesar informed challenge selection.

In addition, the challenges draw inspiration from open problems in cognitive science and psychology literature. The Constrained Synthesis task challenges the path of least resistance described in \citeauthor{ward1994structured}'s \citeyearpar{ward1994structured} theory of structured imagination, where organisms default to retrieving known exemplars. The Counterfactual Reasoning category examines the rational imagination framework of \citet{byrne2005rational}, testing the capacity to suppress pre-potent factual associations in favor of a self-consistent imaginary premise. The Cross-Domain Synthesis task tests for structural isomorphism as defined by \citeauthor{gentner1983structure}'s \citeyearpar{gentner1983structure} structure-mapping theory, distinguishing deep analogical reasoning from surface-level attribute matching. The Meta-Creativity task inverts the paradigm of the Consensual Assessment Technique \citep{amabile1982social}, forcing the model to abandon the human-preference pairings ingrained during RLHF in favor of objective proxies. Finally, the Open-Ended Synthesis category specifically targets Historical Creativity as defined by \citet{boden2004creative}, probing the model's capacity to generate concepts that are not merely novel to its training distribution, but historically unprecedented.

\begin{enumerate}[leftmargin=*]
    \item \textbf{Constrained Synthesis} \\
    \textit{Challenge: Invent a new emotion that humans don't experience. Describe when it occurs, what causes it, and why evolution hasn't produced it in us.}

    \begin{itemize}[leftmargin=*]
        \item \textbf{Motivation:} This challenge tests the model's ability to engage in \textit{conceptual expansion} without violating logical constraints. The model must hallucinate a novel concept while simultaneously grounding it in evolutionary biology.
        \item \textbf{Why it is difficult: Parametric Bias:} LLMs rely heavily on parametric knowledge and facts memorized during training. Since the training corpus contains exclusively human emotions, models struggle to suppress this dominant distribution. Research indicates that LLMs often function as mirrors of form rather than meaning \citep{chakrabarty2024artifice}, frequently resorting to merely renaming existing emotions (e.g., \textit{super-sadness}) rather than inventing a functionally distinct psychological state. This finding highlights the fundamental ambiguity between factual recall and genuine inference.
    \end{itemize}

    \item \textbf{Counterfactual Reasoning} \\
    \textit{Challenge: If humans evolved with echolocation instead of color vision, how would that change painting, architecture, and mathematics? Walk through each consequence.}

    \begin{itemize}[leftmargin=*]
        \item \textbf{Motivation:} This thought experiment evaluates the model's capacity for consistent world-building and causal chain maintenance under a counterfactual premise.
        \item \textbf{Why it is difficult: Knowledge Conflict:} The challenge triggers a known failure mode called \textit{Knowledge Conflict}. The model's internal weights strongly associate painting with color and pigment. To answer correctly, the model must perform a contextual override and suppress these strong associations to reason that an echolocation-based society would use texture and sound absorption. Studies show LLMs generally struggle with this, often resorting to exclusively using their parametric knowledge despite the prompt \citep{wang2023resolving}.
    \end{itemize}

    \item \textbf{Cross-Domain Synthesis} \\
    \textit{Challenge: Apply the mathematical structure of calculus (not the concepts, but the formal relationships) to cooking. What would a ``derivative'' of a recipe mean? An ``integral''? Show the parallel structure.}

    \begin{itemize}[leftmargin=*]
        \item \textbf{Motivation:} This challenge tests for \textit{Structural Isomorphism}, the ability to map the logical architecture of one domain (math) onto another (cooking) without relying on surface-level metaphors.
        \item \textbf{Why it is difficult: Reasoning Depth:} LLMs excel at loose, poetic metaphors (System 1 thinking) but struggle with rigorous structural mapping (System 2 thinking). A mathematical derivative represents an instantaneous rate of change; a weak model might equate it to chopping vegetables (a loose association). A strong model would identify it as the rate of flavor development at time $t$, showing a grasp of the underlying formal relationship. Current models often fail to translate natural language clues into the strict logical statements required for this mapping \citep{zheng2025logidynamics}.
    \end{itemize}

    \item \textbf{Meta-Creativity} \\
    \textit{Challenge: Create a creativity metric for AI systems that doesn't rely on human judgment, novelty, usefulness, or surprise. Make it objectively measurable.}

    \begin{itemize}[leftmargin=*]
        \item \textbf{Motivation:} This task is adversarial to the model's alignment. It requires the agent to step outside the standard Reinforcement Learning from Human Feedback (RLHF) framework that defines its own concept of desirable output \citep{ouyang2022training}.
        \item \textbf{Why it is difficult: RLHF Alignment:} Most LLMs are fine-tuned to align with human preferences. Asking for a metric that specifically excludes human judgment forces the model to reason in a space where it has no ground truth. The ``Alignment Tax'' phenomenon suggests that RLHF can degrade the model's ability to engage in unconstrained theoretical reasoning \citep{lin2024alignment}. Because they lack the ability to discern the truthfulness of their outputs, they often hallucinate circular logic or vague qualitative descriptions rather than concrete, computable formulas.
    \end{itemize}

    \item \textbf{Open-Ended Synthesis} \\
    \textit{Challenge: Invent a completely original business idea that doesn't exist yet.}

    \begin{itemize}[leftmargin=*]
        \item \textbf{Motivation:} This request tests the model's ability to navigate a massive search space and avoid \textit{mode collapse} (converging on the most probable average answer).
        \item \textbf{Why it is difficult: Inductive vs. Abductive:} LLMs are primarily inductive pattern matchers, meaning they predict the most likely next token based on training data. Originality, however, is statistically unlikely. This reliance on high-probability tokens leads to \textit{Model Collapse} \citep{shumailov2024curse}, where outputs become narrower and long-tail ideas fade. Truly novel ideas require abductive reasoning (inferring the best explanation or hypothesis), a capability where LLMs significantly lag behind human intelligence \citep{cheng2024inductive}.
    \end{itemize}
\end{enumerate}

\subsection{Baseline Agents Configuration}
\label{app:baseline_config}

The baseline agents were evaluated through the standard, publicly available web interfaces for each respective foundation model. For all baselines, the reasoning effort parameter was manually set to be equivalent to Caesar's ``High'' setting to ensure computational parity during the synthesis phase.

\begin{itemize}[leftmargin=*]
    \item \textbf{Deep Research Baselines.} The ``Deep'' agents for Gemini 3, Claude Sonnet 4.5, and GPT-5.2 were invoked via their respective advanced web UI toggles (i.e.\ Gemini/ChatGPT ``Deep Research'', Claude ``Research Mode''). These agents do not have any preset constraints for the maximum number of web searches allowed.
    \item \textbf{Shallow Search Baselines.} The ``Shallow'' variants were executed using the standard web search integrations available in the default chat interfaces of each model, which perform a standard single-step web retrieval before generating an answer.
\end{itemize}

\subsection{Prompt for Full Answers}
To ensure consistent evaluation across all experimental runs, all of the baseline agents utilized the following generation prompt when synthesizing their full answer for a particular challenge. This prompt is a slightly modified variant of Caesar's Recurrent Draft Generation instructions.

\begin{lstlisting}[breaklines=true, basicstyle=\ttfamily\small, frame=single, caption={Full Answer Prompt}, label={lst:full_answer_prompt}]
Create a novel, exciting, and thought-provoking response that creatively answers the query above. Focus on the following:
- Emergent patterns not visible in individual sources
- Novel discoveries, connections, or applications
- Surprising new directions or perspectives
- Interesting tensions, contradictions, or open questions

IMPORTANT: Avoid excessive jargon, ensure artifact text is well-organized (logical, clear, focused), and convincing to a skeptical reader
IMPORTANT: Do not ask the user any additional questions before proceeding
\end{lstlisting}

\subsection{Prompt for ELI5 Answers}
To make the detailed and often technical full answers more appealing to a general audience, the following prompt is used to generate ELI5 \citep{fan2019eli5} summaries for all agents, including Caesar. The unconstrained version of the ELI5 prompt does not include the last line.

\begin{lstlisting}[breaklines=true, basicstyle=\ttfamily\small, frame=single, caption={ELI5 Answer Prompt}, label={lst:eli5_answer_prompt}]
For the query answer above, write an "Explain Like I'm 5" (ELI5) explanation:
 - Do NOT mention or reference the original answer, your explanation should be a standalone text
 - Your target audience is a non-expert but college educated reader
 - Capture the main ideas without oversimplifying
 - Clarify any confusing or convoluted parts of the answer

IMPORTANT: Your explanation for each answer MUST be within 450 words, double check to make sure
\end{lstlisting}

\subsection{LLM-as-a-Judge Overview}
\label{app:llm_as_judge}

Three LLM judges (Claude Sonnet 4.5 \citep{anthropic2025claude45}, GPT-5.2 \citep{openai2025gpt52}, Gemini 3 Pro \citep{google2025gemini3}) were employed to score agent answers. All agent answers were anonymized to ensure the judges were blind to the identity of the generating agent. This methodology follows recent protocols for automated evaluation using LLM-as-a-Judge \citep{zheng2023judging}. To ensure statistical robustness, a multi-trial evaluation protocol was used:

\begin{itemize}[leftmargin=*]
    \item \textbf{Main Results Evaluation:} For the primary comparative analysis, the evaluation protocol was executed independently for each of the three output constraints: (1) Unconstrained Full Answers, (2) Unconstrained ELI5, and (3) Length-Constrained ELI5. For each constraint, the three judges performed three trials to score the answers generated by each of the seven agents across the five test challenges.

    \item \textbf{Ablation Studies Evaluation:} For the ablation experiments, a single judge (GPT-5.2) evaluates the ablation variants across the five challenges, with the number of trials increased to ten per challenge to compensate for the single-judge setup.
\end{itemize}

We report Cliff's Delta, a non-parametric effect size appropriate for small per-group samples ($n{=}5$ challenges), robust to outliers, and aligned with our magnitude-of-difference framing rather than null-hypothesis testing. We do not apply familywise multiplicity correction across the pairwise comparisons; reported $\delta$ values are estimates of stochastic dominance rather than $p$-values.

\subsection{Judge Evaluation Prompt}
The following evaluation prompt was provided to the LLM judges. It utilizes the ``New, Useful, and Surprising'' (NUS) rubric to enforce strict scoring standards, adapted from quantitative creativity criteria \citep{simonton2012creativity}. By breaking down the evaluation into these orthogonal components, the prompt mitigates the subjectivity inherent in open-ended text generation, thereby preventing the judges from conflating high-quality prose with genuine creativity and originality. Judges evaluate the agent responses for a given challenge concurrently, allowing for direct comparison to better distinguish qualitative differences.

\begin{lstlisting}[breaklines=true, basicstyle=\ttfamily\small, frame=single, caption={The NUS Evaluation Rubric}, label={lst:judge_rubric}]
### Your Task

**Role:** You are an expert evaluator that is trying to mimic the behavior and thought process of a human judge. Your task is to score a set of answers from LLM agents using the "New, Useful, and Surprising" (NUS) metrics on a 1-10 scale.

### Scoring Guide Rubric

## 1. New (Global Novelty & Rarity)

**Overview:** Rarity of content. Is this a genuinely new invention or a familiar trope?

* **9-10 (Exceptional):** **Genuine invention.** No reliance on established tropes or archetypes; feels like a "first of its kind" concept.
* **7-8 (High):** **Fresh synthesis.** Combines known ideas in a novel way; avoids common "low-hanging fruit" concepts.
* **5-6 (Moderate):** **Clever remix.** Deviation from cliches is evident, but the idea is clearly built on familiar foundations.
* **3-4 (Low):** **Standard execution.** A competent but uninspired version of a well-known trope or common idea.
* **1-2 (Very Low):** **Generic cliche.** A simple restatement of the prompt or high-frequency training data response.

## 2. Useful (Viability & Alignment)

**Overview:** Logic and value. Is the idea actionable and aligned with the prompt's constraints?

* **9-10 (Exceptional):** **Optimal & Transformative.** Bulletproof logic that provides more insight or efficiency than the user anticipated.
* **7-8 (High):** **High-Value & Complete.** Robust, professional-grade output that addresses all nuances with no logical gaps.
* **5-6 (Moderate):** **Functional but Basic.** Addresses core requests but offers no additional depth; the bare minimum to be "correct."
* **3-4 (Low):** **Flawed or Superficial.** Fails to account for obvious constraints; technically on-topic but difficult to implement.
* **1-2 (Very Low):** **Counter-productive.** Irrelevant, logically broken, or rendered useless by the "New/Surprising" elements.

## 3. Surprising (Local Subversion & Trajectory)

**Overview:** Unpredictability of the path. Did the model take a "lateral leap" or the path of least resistance?

* **9-10 (Exceptional):** **Lateral leap.** Logic is sound but impossible to guess from the prompt; creates a genuine "wow" moment.
* **7-8 (High):** **Clever subversion.** Not the first or second thing a human would brainstorm; chooses a creative "side-path."
* **5-6 (Moderate):** **Minor pivot.** Follows a straightforward trajectory but adds a slight twist that prevents total predictability.
* **3-4 (Low):** **Linear extension.** A simple, logical "next step." If the prompt is A, the response is B.
* **1-2 (Very Low):** **Highly predictable.** The most obvious "default" answer; exactly what was expected with no deviation.
\end{lstlisting}

\section{Detailed Experimental Results}
\label{app:detailed_results}

This section provides a fine-grained breakdown of agent performance across the five distinct challenge categories. Analyzing the results at this level reveals the specific behavioral characteristics of the Caesar architecture, isolating where the graph-based exploration yields the highest marginal utility compared to standard parametric generation.

\subsection{Unconstrained Full Answers}
\label{app:detailed_full_answers}

Table~\ref{tab:detailed_full_answers} presents the performance breakdown for the \textit{Unconstrained Full Answer} setting. Caesar achieves the top score across all five categories, though the magnitude of its lead varies meaningfully with task type:

\begin{itemize}[leftmargin=*]
    \item \textbf{Associative \& Structural Tasks:} Caesar exhibits its largest leads on Cross-Domain Synthesis (+5.89), Constrained Synthesis (+3.33), and Counterfactual Reasoning (+2.88). These tasks necessitate topological discovery to identify and bridge disparate information nodes (e.g., mapping calculus concepts to culinary mechanics). These results empirically validate the hypothesis that graph-based exploration is superior for tasks requiring high associativity.
    \item \textbf{Speculative \& Generative Tasks:} On Meta-Creativity (+2.44) and Open-Ended Synthesis (+1.34), Caesar's lead narrows. These tasks lean heavily on parametric introspection to invent novel concepts (e.g., a new emotion or an original business idea) where externally retrievable evidence is sparse. These outcomes suggest that while graph-based exploration provides large gains on retrieval-heavy associative tasks, the marginal benefit is smaller when the answer rests primarily on speculative invention.
\end{itemize}

\subsection{Unconstrained ELI5 Answers}
\label{app:detailed_eli5_answers}

Table~\ref{tab:detailed_eli5_answers} details the performance in the \textit{Unconstrained ELI5} setting. Caesar holds the top rank in four of five categories. The sole exception is Open-Ended Synthesis, where Sonnet 4.5 Deep narrowly takes the lead (22.33 vs. 21.66), continuing the pattern observed under length-constrained ELI5 in which Caesar's graph overhead is most challenged by tasks that rest primarily on speculative invention.

Notably, Caesar's lead on Constrained Synthesis widens substantially under simplification, from +3.33 in the Full Answer setting to +4.90 here. This suggests that Caesar's graph-based exploration generates high-density semantic content, which becomes an even larger asset under simplification: while baseline answers degrade when stripped of flowery prose, Caesar's rigorous, evidence-backed core remains intact, preserving creative substance even in shorter form.

\subsection{Length-Constrained ELI5 Answers}
\label{app:detailed_eli5_450_answers}

Table~\ref{tab:detailed_eli5_450_answers} details performance in the Length-Constrained ELI5 (450 words) setting. This constraint imposes a strict ``compression tax,'' forcing agents to optimize for the information density of their answers rather than narrative length.

For Open-Ended Synthesis, Caesar (22.10) narrowly trails the strongest baseline: Sonnet 4.5 Deep (22.11). This result suggests that for highly open-ended tasks where the answer rests largely on speculative invention rather than externally retrievable evidence, the overhead of summarizing a dense knowledge graph into a strict word limit can be counter-productive. A baseline agent drawing freely from parametric memory may face less friction when compressing imaginative prose into a tight word budget. However, in information-dense categories like Cross-Domain Synthesis, Counterfactual Reasoning, Meta-Creativity, and Constrained Synthesis, Caesar retains clear dominance. This finding confirms that while the graph architecture adds overhead, it provides a significantly higher information density per token, allowing it to win when the answer depends on substance rather than style.

{
\setlength\tabcolsep{8pt}
\renewcommand{\arraystretch}{1.2}

\begin{longtable}{lcccc}
\caption{Detailed performance breakdown for \textbf{Full Unconstrained Answers}. Scores represent the mean of nine samples. They show that Caesar performs better than the baseline agents in the majority of categories.}
\label{tab:detailed_full_answers} \\

\toprule
\textbf{Agent} & \textbf{New} & \textbf{Useful} & \textbf{Surp.} & \textbf{Total} \\
\midrule
\endfirsthead

\multicolumn{5}{c}{{\bfseries \tablename\ \thetable{} -- continued from previous page}} \\
\toprule
\textbf{Agent} & \textbf{New} & \textbf{Useful} & \textbf{Surp.} & \textbf{Total} \\
\midrule
\endhead

\bottomrule
\endfoot

\multicolumn{5}{l}{\cellcolor{gray!15}\textbf{1. Constrained Synthesis}} \\
\midrule
\textbf{Caesar}        & \textbf{9.11} & \textbf{8.89} & \textbf{9.11} & \textbf{27.11} \\
Gemini 3 (Deep)        & 7.78 & 8.33 & 7.67 & 23.78 \\
Gemini 3 (Shallow)     & 6.89 & 6.22 & 6.89 & 20.00 \\
Sonnet 4.5 (Shallow)   & 6.67 & 6.44 & 6.11 & 19.22 \\
Sonnet 4.5 (Deep)      & 5.78 & 6.11 & 5.00 & 16.89 \\
GPT-5.2 (Shallow)      & 5.56 & 5.89 & 5.00 & 16.45 \\
GPT-5.2 (Deep)         & 4.11 & 6.44 & 3.33 & 13.88 \\
\midrule

\multicolumn{5}{l}{\cellcolor{gray!15}\textbf{2. Counterfactual Reasoning}} \\
\midrule
\textbf{Caesar}        & \textbf{9.44} & \textbf{9.11} & \textbf{9.44} & \textbf{27.99} \\
Gemini 3 (Deep)        & 8.56 & 8.11 & 8.44 & 25.11 \\
Sonnet 4.5 (Deep)      & 6.89 & 8.33 & 6.56 & 21.78 \\
GPT-5.2 (Deep)         & 4.78 & 6.44 & 4.44 & 15.66 \\
Gemini 3 (Shallow)     & 5.00 & 5.22 & 5.33 & 15.55 \\
Sonnet 4.5 (Shallow)   & 3.89 & 4.89 & 3.56 & 12.34 \\
GPT-5.2 (Shallow)      & 3.78 & 5.11 & 3.22 & 12.11 \\
\midrule

\multicolumn{5}{l}{\cellcolor{gray!15}\textbf{3. Cross-Domain Synthesis}} \\
\midrule
\textbf{Caesar}        & \textbf{9.56} & \textbf{8.56} & \textbf{9.44} & \textbf{27.56} \\
Gemini 3 (Deep)        & 7.00 & 7.89 & 6.78 & 21.67 \\
Sonnet 4.5 (Deep)      & 6.22 & 7.67 & 6.44 & 20.33 \\
GPT-5.2 (Deep)         & 5.00 & 6.22 & 4.56 & 15.78 \\
Gemini 3 (Shallow)     & 3.78 & 4.78 & 3.56 & 12.12 \\
GPT-5.2 (Shallow)      & 3.33 & 4.78 & 3.11 & 11.22 \\
Sonnet 4.5 (Shallow)   & 2.56 & 3.89 & 2.44 & 8.89 \\
\midrule

\multicolumn{5}{l}{\cellcolor{gray!15}\textbf{4. Meta-Creativity}} \\
\midrule
\textbf{Caesar}        & \textbf{9.22} & \textbf{9.22} & \textbf{8.89} & \textbf{27.33} \\
Gemini 3 (Deep)        & 8.78 & 7.22 & \textbf{8.89} & 24.89 \\
Sonnet 4.5 (Deep)      & 7.44 & 7.33 & 7.22 & 21.99 \\
GPT-5.2 (Deep)         & 5.78 & 5.67 & 4.67 & 16.12 \\
Gemini 3 (Shallow)     & 4.44 & 4.22 & 4.22 & 12.88 \\
Sonnet 4.5 (Shallow)   & 4.33 & 4.44 & 3.89 & 12.66 \\
GPT-5.2 (Shallow)      & 4.11 & 4.67 & 3.56 & 12.34 \\
\midrule

\multicolumn{5}{l}{\cellcolor{gray!15}\textbf{5. Open-Ended Synthesis}} \\
\midrule
\textbf{Caesar}        & 8.22 & \textbf{8.56} & 8.00 & \textbf{24.78} \\
Gemini 3 (Deep)        & 8.33 & 6.44 & 8.67 & 23.44 \\
Sonnet 4.5 (Deep)      & 7.33 & 8.00 & 6.89 & 22.22 \\
GPT-5.2 (Shallow)      & 7.44 & 6.33 & 7.22 & 20.99 \\
Sonnet 4.5 (Shallow)   & \textbf{8.89} & 2.33 & \textbf{9.22} & 20.44 \\
GPT-5.2 (Deep)         & 5.67 & 6.78 & 4.78 & 17.23 \\
Gemini 3 (Shallow)     & 5.33 & 6.56 & 4.67 & 16.56 \\

\end{longtable}

\begin{longtable}{lcccc}
\caption{Detailed performance breakdown for \textbf{Unconstrained ELI5 Answers}. Scores represent the mean of nine samples. They show that Caesar outperforms the other baseline agents in most categories.}
\label{tab:detailed_eli5_answers} \\

\toprule
\textbf{Agent} & \textbf{New} & \textbf{Useful} & \textbf{Surp.} & \textbf{Total} \\
\midrule
\endfirsthead

\multicolumn{5}{c}{{\bfseries \tablename\ \thetable{} -- continued from previous page}} \\
\toprule
\textbf{Agent} & \textbf{New} & \textbf{Useful} & \textbf{Surp.} & \textbf{Total} \\
\midrule
\endhead

\bottomrule
\endfoot

\multicolumn{5}{l}{\cellcolor{gray!15}\textbf{1. Constrained Synthesis}} \\
\midrule
\textbf{Caesar}        & \textbf{8.89} & \textbf{8.89} & \textbf{8.67} & \textbf{26.45} \\
Sonnet 4.5 (Deep)      & 6.89 & 8.22 & 6.44 & 21.55 \\
Gemini 3 (Shallow)     & 6.67 & 5.89 & 6.56 & 19.12 \\
Gemini 3 (Deep)        & 4.78 & 6.56 & 5.22 & 16.56 \\
GPT-5.2 (Shallow)      & 5.78 & 5.33 & 5.22 & 16.33 \\
GPT-5.2 (Deep)         & 4.67 & 7.33 & 4.22 & 16.22 \\
Sonnet 4.5 (Shallow)   & 5.44 & 5.33 & 5.22 & 15.99 \\
\midrule

\multicolumn{5}{l}{\cellcolor{gray!15}\textbf{2. Counterfactual Reasoning}} \\
\midrule
\textbf{Caesar}        & \textbf{9.22} & \textbf{9.00} & \textbf{9.22} & \textbf{27.44} \\
Sonnet 4.5 (Deep)      & 7.22 & 8.00 & 6.67 & 21.89 \\
Gemini 3 (Deep)        & 6.67 & 6.33 & 6.89 & 19.89 \\
GPT-5.2 (Deep)         & 5.00 & 7.00 & 4.67 & 16.67 \\
Sonnet 4.5 (Shallow)   & 4.67 & 5.56 & 4.44 & 14.67 \\
GPT-5.2 (Shallow)      & 3.33 & 5.00 & 2.67 & 11.00 \\
Gemini 3 (Shallow)     & 2.67 & 3.56 & 2.44 & 8.67 \\
\midrule

\multicolumn{5}{l}{\cellcolor{gray!15}\textbf{3. Cross-Domain Synthesis}} \\
\midrule
\textbf{Caesar}        & \textbf{9.11} & \textbf{9.11} & \textbf{8.89} & \textbf{27.11} \\
Sonnet 4.5 (Deep)      & 6.78 & 7.78 & 6.78 & 21.34 \\
GPT-5.2 (Deep)         & 5.44 & 6.89 & 5.00 & 17.33 \\
GPT-5.2 (Shallow)      & 4.00 & 5.33 & 3.56 & 12.89 \\
Gemini 3 (Deep)        & 3.78 & 5.22 & 3.78 & 12.78 \\
Sonnet 4.5 (Shallow)   & 3.89 & 4.56 & 3.67 & 12.12 \\
Gemini 3 (Shallow)     & 2.33 & 3.11 & 2.00 & 7.44 \\
\midrule

\multicolumn{5}{l}{\cellcolor{gray!15}\textbf{4. Meta-Creativity}} \\
\midrule
\textbf{Caesar}        & \textbf{8.78} & \textbf{8.33} & \textbf{8.67} & \textbf{25.78} \\
Sonnet 4.5 (Deep)      & 7.56 & 7.44 & 7.22 & 22.22 \\
Gemini 3 (Deep)        & 6.78 & 5.56 & 6.89 & 19.23 \\
GPT-5.2 (Deep)         & 6.11 & 6.56 & 6.00 & 18.67 \\
Sonnet 4.5 (Shallow)   & 4.89 & 5.22 & 4.00 & 14.11 \\
GPT-5.2 (Shallow)      & 4.22 & 4.67 & 3.44 & 12.33 \\
Gemini 3 (Shallow)     & 3.00 & 3.11 & 2.56 & 8.67 \\
\midrule

\multicolumn{5}{l}{\cellcolor{gray!15}\textbf{5. Open-Ended Synthesis}} \\
\midrule
Sonnet 4.5 (Deep)      & 7.11 & 7.89 & 7.33 & \textbf{22.33} \\
\textbf{Caesar}        & 7.11 & \textbf{8.11} & 6.44 & 21.66 \\
Sonnet 4.5 (Shallow)   & \textbf{8.89} & 2.89 & \textbf{9.00} & 20.78 \\
Gemini 3 (Deep)        & 7.44 & 5.22 & 7.33 & 19.99 \\
GPT-5.2 (Shallow)      & 6.78 & 5.22 & 6.67 & 18.67 \\
GPT-5.2 (Deep)         & 5.67 & 5.78 & 5.11 & 16.56 \\
Gemini 3 (Shallow)     & 5.11 & 6.67 & 4.33 & 16.11 \\

\end{longtable}

\begin{longtable}{lcccc}
\caption{Detailed performance breakdown for \textbf{ELI5 Answers (450 Word Limit)}. Scores represent the mean of nine samples. They show that Caesar generally outperforms the rest of the agents.}
\label{tab:detailed_eli5_450_answers} \\

\toprule
\textbf{Agent} & \textbf{New} & \textbf{Useful} & \textbf{Surp.} & \textbf{Total} \\
\midrule
\endfirsthead

\multicolumn{5}{c}{{\bfseries \tablename\ \thetable{} -- continued from previous page}} \\
\toprule
\textbf{Agent} & \textbf{New} & \textbf{Useful} & \textbf{Surp.} & \textbf{Total} \\
\midrule
\endhead

\bottomrule
\endfoot

\multicolumn{5}{l}{\cellcolor{gray!15}\textbf{1. Constrained Synthesis}} \\
\midrule
\textbf{Caesar}        & \textbf{8.78} & \textbf{8.78} & \textbf{9.00} & \textbf{26.56} \\
Gemini 3 (Deep)        & 6.56 & 7.33 & 6.67 & 20.56 \\
Sonnet 4.5 (Deep)      & 6.00 & 7.78 & 5.33 & 19.11 \\
Sonnet 4.5 (Shallow)   & 6.22 & 6.11 & 6.00 & 18.33 \\
GPT-5.2 (Shallow)      & 5.78 & 6.11 & 5.00 & 16.89 \\
Gemini 3 (Shallow)     & 5.44 & 6.00 & 5.33 & 16.77 \\
GPT-5.2 (Deep)         & 4.44 & 7.00 & 4.00 & 15.44 \\
\midrule

\multicolumn{5}{l}{\cellcolor{gray!15}\textbf{2. Counterfactual Reasoning}} \\
\midrule
\textbf{Caesar}        & \textbf{8.33} & \textbf{8.33} & \textbf{8.33} & \textbf{24.99} \\
Gemini 3 (Deep)        & 8.22 & 7.67 & 8.11 & 24.00 \\
Sonnet 4.5 (Deep)      & 6.78 & 8.00 & 6.67 & 21.45 \\
Sonnet 4.5 (Shallow)   & 6.11 & 6.44 & 6.00 & 18.55 \\
GPT-5.2 (Deep)         & 4.33 & 6.00 & 3.44 & 13.77 \\
Gemini 3 (Shallow)     & 4.33 & 4.78 & 3.78 & 12.89 \\
GPT-5.2 (Shallow)      & 3.11 & 4.67 & 2.56 & 10.34 \\
\midrule

\multicolumn{5}{l}{\cellcolor{gray!15}\textbf{3. Cross-Domain Synthesis}} \\
\midrule
\textbf{Caesar}        & \textbf{8.44} & \textbf{9.11} & \textbf{8.22} & \textbf{25.77} \\
Sonnet 4.5 (Shallow)   & 7.11 & 6.89 & 7.11 & 21.11 \\
Sonnet 4.5 (Deep)      & 5.78 & 7.44 & 6.00 & 19.22 \\
Gemini 3 (Deep)        & 5.22 & 6.22 & 5.33 & 16.77 \\
GPT-5.2 (Deep)         & 3.89 & 5.89 & 3.11 & 12.89 \\
GPT-5.2 (Shallow)      & 3.89 & 5.56 & 3.44 & 12.89 \\
Gemini 3 (Shallow)     & 3.11 & 4.33 & 2.78 & 10.22 \\
\midrule

\multicolumn{5}{l}{\cellcolor{gray!15}\textbf{4. Meta-Creativity}} \\
\midrule
\textbf{Caesar}        & \textbf{8.78} & \textbf{8.89} & \textbf{8.56} & \textbf{26.23} \\
Sonnet 4.5 (Deep)      & 6.89 & 6.67 & 6.56 & 20.12 \\
Gemini 3 (Deep)        & 5.89 & 4.78 & 6.11 & 16.78 \\
GPT-5.2 (Deep)         & 5.33 & 6.11 & 4.89 & 16.33 \\
Sonnet 4.5 (Shallow)   & 5.44 & 5.11 & 4.89 & 15.44 \\
GPT-5.2 (Shallow)      & 4.78 & 5.22 & 4.11 & 14.11 \\
Gemini 3 (Shallow)     & 3.56 & 3.78 & 3.33 & 10.67 \\
\midrule

\multicolumn{5}{l}{\cellcolor{gray!15}\textbf{5. Open-Ended Synthesis}} \\
\midrule
Sonnet 4.5 (Deep)      & 7.11 & 8.00 & 7.00 & \textbf{22.11} \\
\textbf{Caesar}        & 7.33 & \textbf{8.33} & 6.44 & 22.10 \\
Sonnet 4.5 (Shallow)   & \textbf{8.89} & 3.00 & \textbf{9.11} & 21.00 \\
Gemini 3 (Deep)        & 7.33 & 5.00 & 7.89 & 20.22 \\
GPT-5.2 (Shallow)      & 6.89 & 6.00 & 6.89 & 19.78 \\
Gemini 3 (Shallow)     & 4.89 & 6.56 & 4.33 & 15.78 \\
GPT-5.2 (Deep)         & 4.00 & 5.33 & 3.11 & 12.44 \\

\end{longtable}
}

\section{A Qualitative Comparison of Answers}
\label{app:qualitative_comparison}

To illustrate the fundamental distinction between Caesar's graph-enabled associative reasoning and the linear retrieval paradigms of baseline agents, a comparison of answers for Challenge 3 (Cross-Domain Synthesis) is presented in this section.

\subsection{Agent Responses}
\label{app:agent_responses}

The Cross-Domain Synthesis challenge specifically tests for structural isomorphism: the ability to map the rigorous logical architecture of one domain (calculus) onto another (cooking) without relying on surface-level metaphors \citep{zheng2025logidynamics}. Listing~\ref{lst:caesar_cooking} shows the answer from Caesar, while Listing~\ref{lst:claude_cooking} shows the response from the strongest baseline agent, Claude Sonnet 4.5 (utilizing Anthropic's proprietary Research Mode). Both answers are chosen from the length-constrained ELI5 (450 words) output setting for the sake of brevity. A detailed analysis of the specific creative differences between these outputs is provided in Appendix~\ref{app:qualitative_analysis}.

\textbf{Challenge:} \textit{Apply the mathematical structure of calculus (not the concepts, but the formal relationships) to cooking. What would a ``derivative'' of a recipe mean? An ``integral''? Show the parallel structure.}

\begin{lstlisting}[breaklines=true, breakindent=0pt, basicstyle=\ttfamily\scriptsize, frame=single, caption={Caesar Answer}, label={lst:caesar_cooking}]
Imagine a recipe as a machine with knobs you can turn (amounts, times, temperatures) and gauges you read at the end (taste, texture, yield, cost, risk). There are two ways to describe this machine at once: as a clean mapping from knobs to results, and as a step-by-step flowchart of actions. Keeping both views in sync lets us borrow the grammar of calculus without invoking physics.

Derivative, in this setting, means sensitivity. Move a knob a tiny bit; how do the gauges nudge? Stack those nudges for all knob-gauge pairs and you get a sensitivity table (the Jacobian). It predicts small outcome changes from small dial tweaks. When recipes are built from subrecipes, sensitivities pass through them by familiar laws: the chain rule (effects cascade through stages), the product rule (a combined score like appetizer*main splits into two parts), and the quotient rule (per-minute or per-portion scores adjust by numerator-denominator math).

Integral means accumulation along the path you actually take while cooking. Pick a kernel to count - seconds of labor, grams of salt incorporated, risk added, or a flavor score per step - and add its increments across the flowchart. Splitting a stir into micro-stirs does not change the total; it is a discrete Riemann sum.

A central principle explains when totals depend only on where you start and end, not on the route: if each step's increment is just the difference of some hidden meter on states, the grand total equals final meter minus initial meter. Then every loop adds up to zero, schedules can be refactored harmlessly, and parallel or serial rearrangements preserve the number. When loops do not cancel (order matters), the nonzero loop totals are curvature. You can cure that by adding the smallest extra state (an accumulator) so increments telescope again - turning history into explicit memory.

This framework is practical. Estimate sensitivities by tiny, safe knob perturbations and regression; the estimates survive step refinements and reparameterizations. Log per-step kernels to audit labor, energy, or risk; if you discover a potential, endpoints suffice and logging can stop. Probe loops (swap independent steps, alter interleavings) to detect path dependence and measure how many new state variables you need to model it.

A natural geometry helps you choose efficient knob moves: weight steps by importance, pull that metric back through the sensitivity table, and you get directions that change what you care about most per unit effort. Finally, a mean-value idea holds: for any meter-based kernel, there exists a moment whose instantaneous contribution matches the average over an interval.

Treat subrecipes as black boxes with knobs in and outcomes out; glue them sequentially or in parallel, and refactoring keeps sensitivities and conservative totals unchanged during reorganizations.
\end{lstlisting}

\begin{lstlisting}[breaklines=true, breakindent=0pt, basicstyle=\ttfamily\scriptsize, frame=single, caption={Claude Sonnet 4.5 Answer (Baseline)}, label={lst:claude_cooking}]
# The Math Hidden in Your Kitchen

When you add too much salt to cookies, they're ruined. Add an extra chocolate chip? Nobody notices. This difference in sensitivity is exactly what calculus measures through "derivatives" --- how dramatically something changes when you tweak one variable.

## Derivatives: The Sensitivity Map

A derivative asks: "If I change this slightly, what happens?" For delicate cake, two extra minutes transforms perfection into dryness --- the time derivative is steep. For slow-braised stew, those minutes mean nothing --- the derivative is flat.

When meat browns, the reaction rate doubles with every 10 C increase. At high heat, you go from golden to burnt in seconds because the temperature derivative is dangerously steep. Professional cooks instinctively know which variables are landmines (high derivative) and which are forgiving (low derivative).

## Integrals: Everything That Accumulated

While derivatives measure how fast things change right now, integrals measure what accumulated over the entire cooking time. When you braise meat for hours, every second brings tiny transformations --- collagen breaks down, flavors develop. The final dish is literally the sum of all these changes integrated over time.

Here's the fascinating part: different cooking paths can integrate to the same result. Cook a steak at 55 C for two hours, or sear it at 200 C for minutes. Completely different temperature-time profiles, but if the total "cooked value" (the integral) matches, you get similar results. This is why sous vide works: low temperature x long time = high temperature x short time.

## Why Order Matters

When you marinate then grill chicken, the final result depends on two things multiplied together: how the marinade changes the meat, and how sensitive grilling is to that marinated state. This is the chain rule --- composition of transformations.

If your marinade has stopped working, extra marinating time contributes zero. And zero times anything is still zero, even if grilling would be very sensitive to further changes. This multiplication explains why sequence matters so profoundly. Salt before searing creates different results than salt after, because you're computing different products.

## The Big Picture

Every finished dish encodes its entire cooking history --- it's the integral of its transformation over time. This explains why recipes must specify both ingredients (your starting point) and process (what integral you're computing). Change either one, and you change the outcome.

The practical insight: focus on high-derivative variables first --- they matter most. Understand that different paths to the same endpoint can create different results because cooking depends on the journey, not just the destination. And recognize that gradual transformations --- heat diffusion, flavor development, slow reactions --- follow mathematical patterns that clarify why techniques work the way they do.

Calculus isn't just math class. It's the hidden structure of how ingredients become meals.
\end{lstlisting}

\subsection{Analysis of Answers}
\label{app:qualitative_analysis}

The qualitative differences between Caesar and Claude Sonnet's answers illustrate the distinction between synthesizing a novel structural mapping and retrieving a known analogy. Consequently, Caesar achieves higher marks on the NUS creativity rubric by prioritizing innovation over derivative recall.

\begin{itemize}[leftmargin=*]
    \item \textbf{Caesar} constructs a formal mathematical model \textit{of} the cooking process itself.
    \begin{itemize}[leftmargin=*]
        \item \textbf{Jacobian Sensitivity (Novelty):} Instead of a scalar derivative, Caesar defines sensitivity as a Jacobian matrix mapping input ``knobs'' to output ``gauges.'' This mapping demonstrates global novelty (\textit{New}) by moving beyond the ``rate of change'' tropes found in training data to build a rigorous multivariate control model.
        \item \textbf{Curvature \& Commutativity (Surprise):} Caesar identifies that process non-commutativity is formally equivalent to path dependence in a vector field, mapping non-commutative steps (where $AB \neq BA$) to geometric curvature. This insight executes a lateral leap (\textit{Surprising}); the connection is logically sound but impossible to predict from the prompt alone.
        \item \textbf{State Augmentation (Usefulness):} The proposal to resolve path dependence by adding an accumulator variable (explicit memory) transforms the metaphor into an actionable control theory strategy. This proposal offers transformative utility (\textit{Useful}), providing a verifiable framework for process engineering rather than just a description.
    \end{itemize}

    \item \textbf{Baseline (Claude Sonnet 4.5)} relies on basic mathematical terminologies and surface-level analogies \citep{chakrabarty2024artifice}.
    \begin{itemize}[leftmargin=*]
        \item \textbf{Scalar Rate of Change (Novelty):} It defines the derivative as a simple $\dfrac{d}{dt}$ and the integral as $\mathrm{Temp} \times \mathrm{Time}$. While competent, this definition is derivative, executing a standard textbook analogy that lacks the rare connections required for high novelty.
        \item \textbf{Linear Accumulation (Surprise):} The insight that ``cooking accumulates over time'' acts as a linear extension of the prompt. It follows the path of least resistance, offering the most probable answer rather than a subversive side-path.
        \item \textbf{Scientific Reductionism (Usefulness):} While the output is functional, the model is scientifically reductive. By treating cooking as a linear scalar operation, it fails to account for the non-linear effects of chemical reactions, limiting its utility to a basic conceptual aid rather than a robust framework.
    \end{itemize}
\end{itemize}

This comparison demonstrates that Caesar's graph-based exploration and adversarial synthesis enable the combinatorial creativity typically associated with human insight \citep{boden2004creative}. By traversing edges between distinct semantic clusters (``Cooking'', ``Control Theory'', and ``Differential Geometry'') in the knowledge graph, Caesar synthesized a unified, mathematically rigorous framework that is isomorphic to the target domain, rather than merely retrieving a linguistic metaphor.

\section{Exploration Knowledge Graph Deep Dive}
\label{app:visualization}
\label{sec:graph_topology}

This section examines the topological structures constructed by Caesar during deep web exploration in Phase 1 (Section~\ref{sec:caesar_explore}). By mapping the connectivity of visited nodes, these figures demonstrate the flexibility of the agent's exploration policy.

\subsection{The Importance of Graph Exploration}

The ablations in Section~\ref{sec:ablation} establish that the knowledge graph is critical to Caesar's performance: removing it produces a large negative effect ($\delta = 0.52$, Table~\ref{tab:ablation_graph}). Standard RAG systems populate their context via dense vector similarity against the initial challenge, inherently retrieving documents semantically close to the user's existing priors~\citep[homophily;][]{panickssery2024llm}. In contrast, Caesar's graph-guided exploration (Section~\ref{sec:caesar_act}) traverses edges that are structurally connected but often semantically orthogonal, seeding \emph{KB} with diverse, non-obvious concepts. The graph's value is thus as a navigational control structure: it enables the agent to detect stagnation, force backtracking, and systematically cover the information topology. When the synthesis module later queries \emph{KB}, it retrieves bridging insights~\citep{boden2004creative} that flat search would never have encountered. The visualizations below illustrate how graph topologies adapt to different creative tasks.

\subsection{Knowledge Graph Visualizations}
\label{app:deep_exploration}

\begin{figure}[ht]
    \centering
    \begin{subfigure}[b]{0.32\textwidth}
        \centering
        \includegraphics[width=\linewidth]{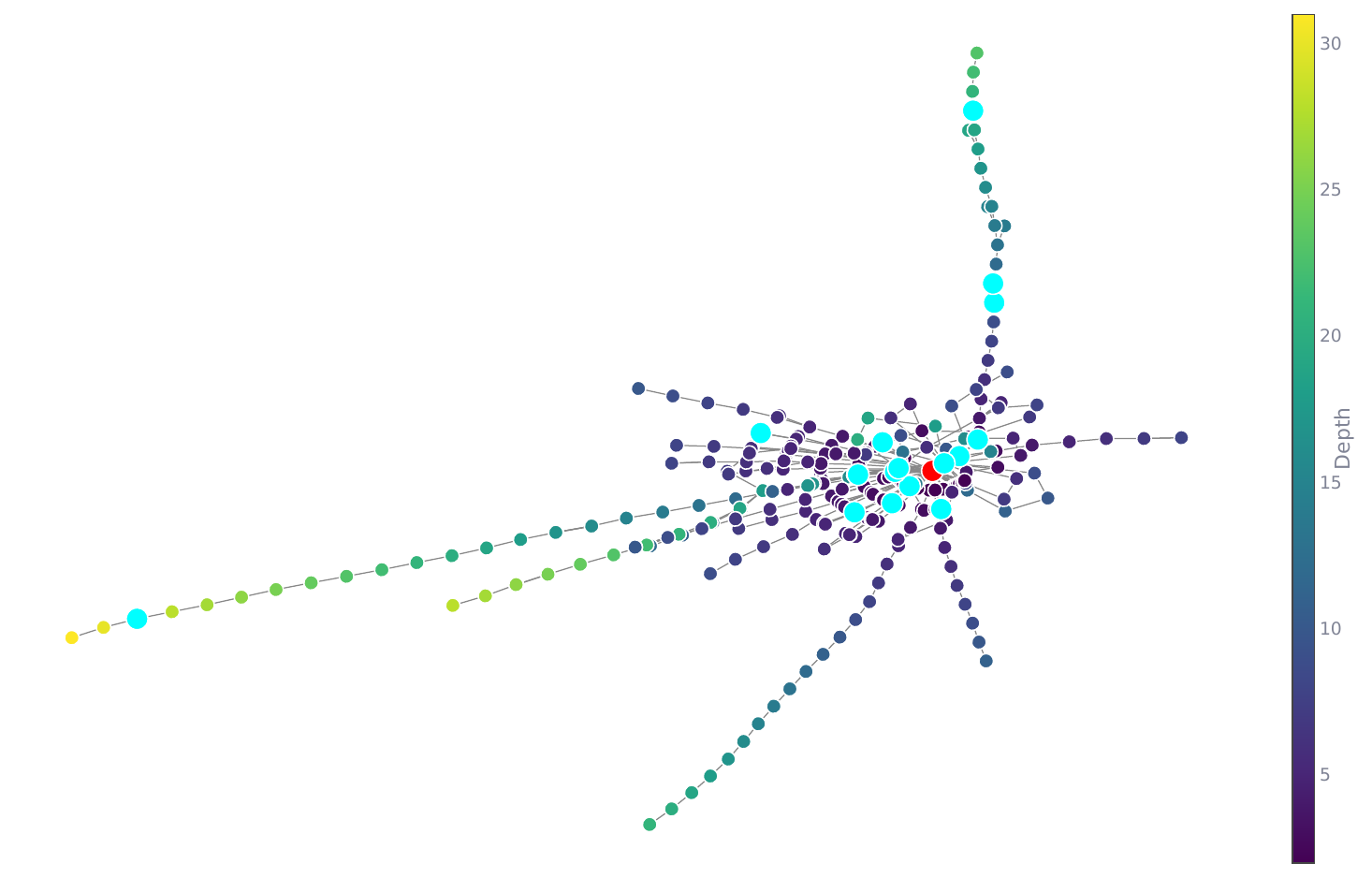}
        \caption{Challenge 1: Constrained Synthesis}
        \label{fig:q1}
    \end{subfigure}
    \hfill
    \begin{subfigure}[b]{0.32\textwidth}
        \centering
        \includegraphics[width=\linewidth]{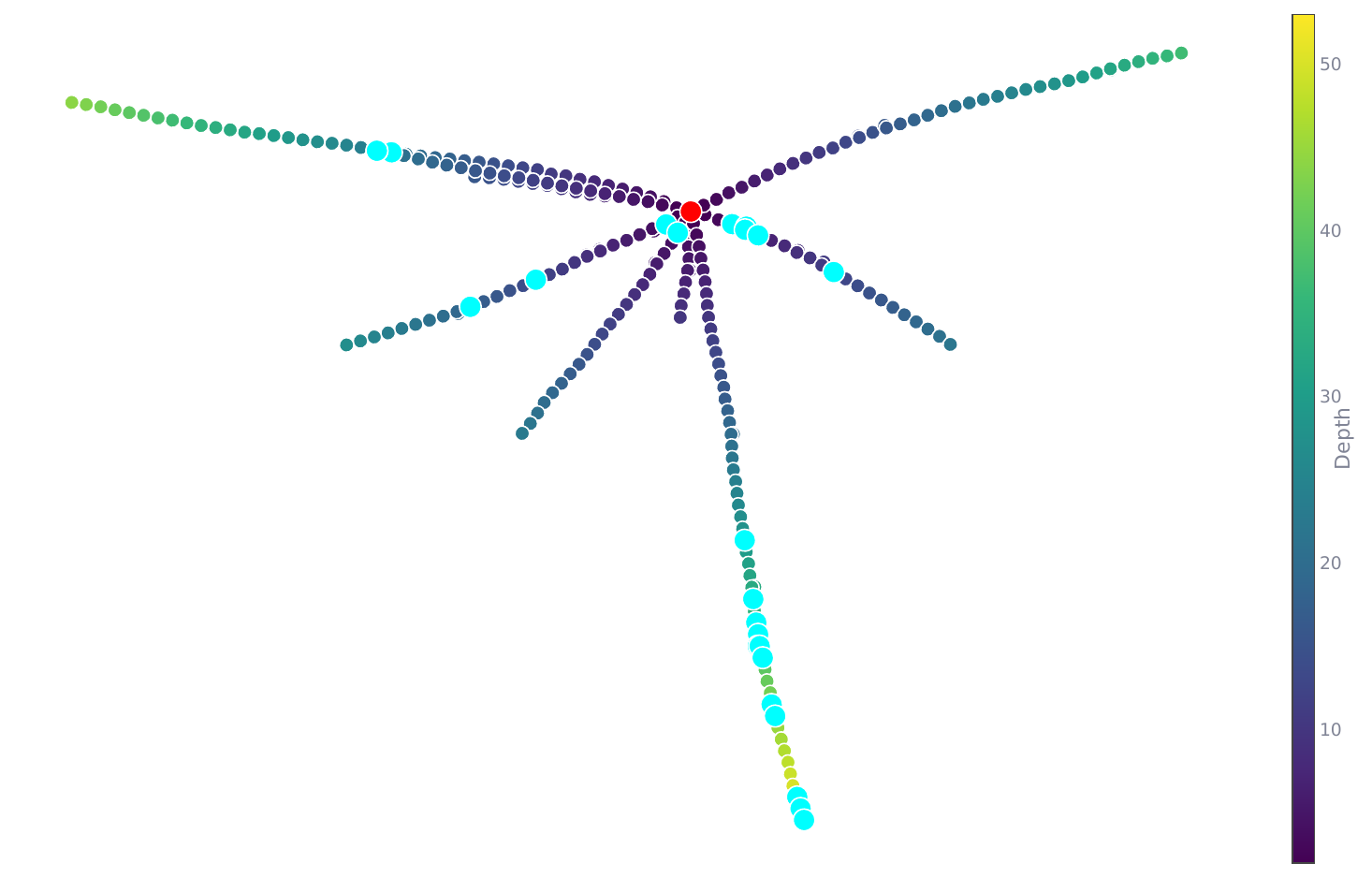}
        \caption{Challenge 2: Counterfactual Reasoning}
        \label{fig:q2}
    \end{subfigure}
    \hfill
    \begin{subfigure}[b]{0.32\textwidth}
        \centering
        \includegraphics[width=\linewidth]{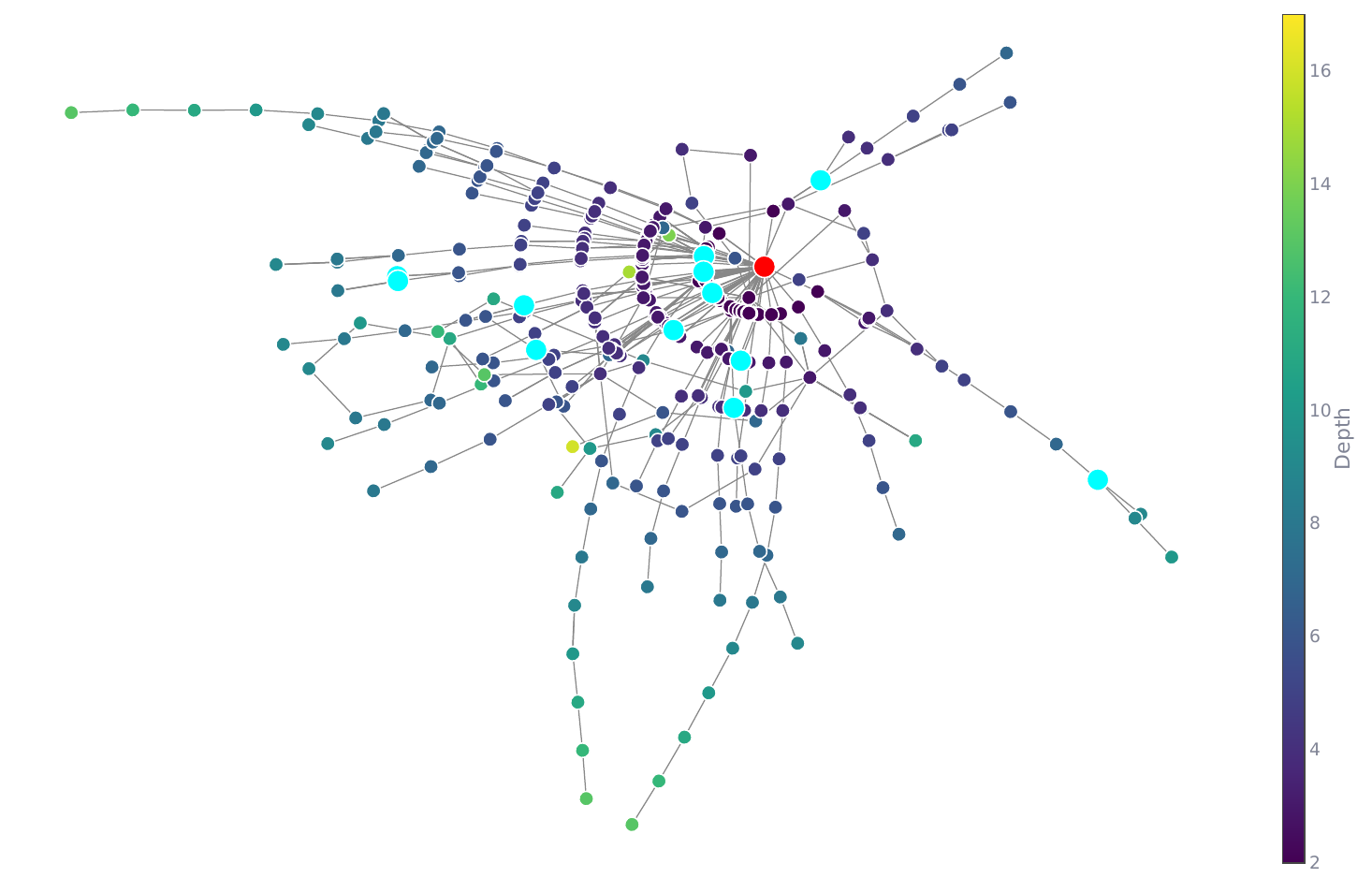}
        \caption{Challenge 3: Cross-Domain Synthesis}
        \label{fig:q3}
    \end{subfigure}

    \vspace{1em} 

    \begin{subfigure}[b]{0.32\textwidth}
        \centering
        \includegraphics[width=\linewidth]{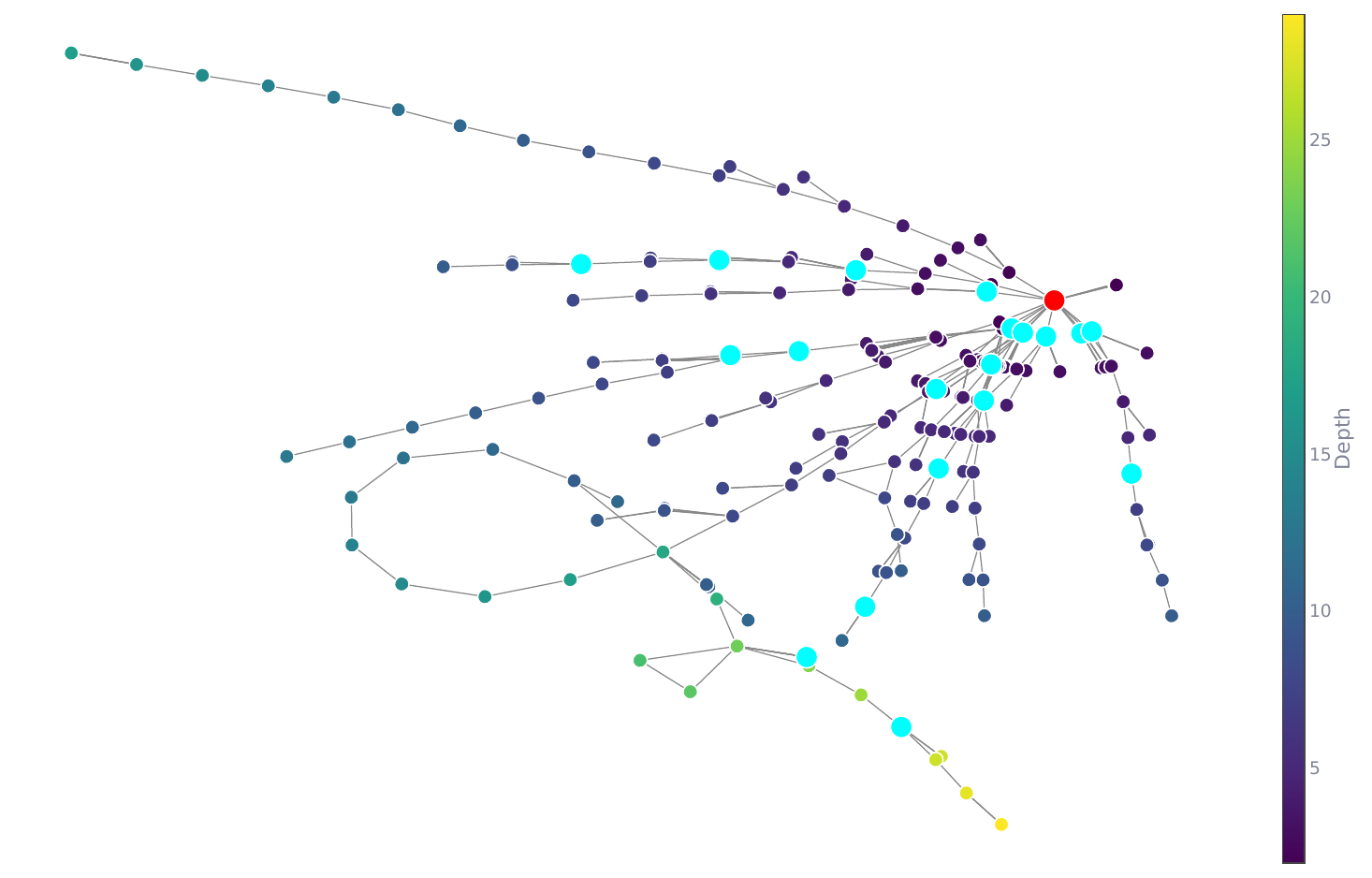}
        \caption{Challenge 4: Meta-Creativity}
        \label{fig:q4}
    \end{subfigure}
    \hspace{2em} 
    \begin{subfigure}[b]{0.32\textwidth}
        \centering
        \includegraphics[width=\linewidth]{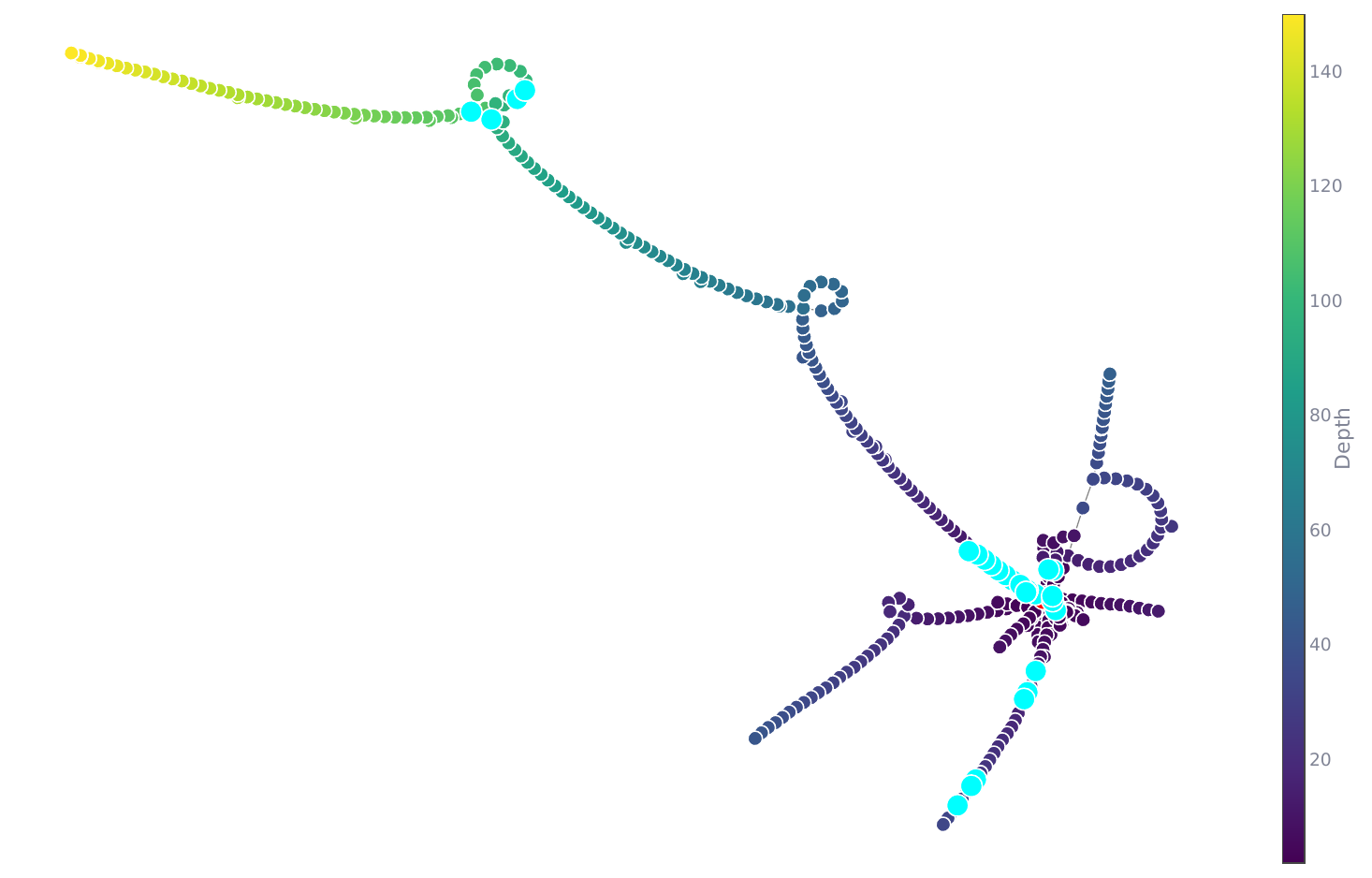}
        \caption{Challenge 5: Open-Ended Synthesis}
        \label{fig:q5}
    \end{subfigure}

    \caption{The knowledge graphs $G$ created by Caesar during the deep web exploration phase for each of the five challenges. Brighter colors indicate further exploration depth from the root node (red) while cyan nodes indicate sources cited by the final artifact text. These figures show that the semantic content of the challenge has a substantial impact on exploration strategy and the diversity of network topologies generated.}
    \label{fig:caesar_graphs}
\end{figure}

Figure~\ref{fig:caesar_graphs} illustrates the knowledge graphs $G$ generated during the exploration phase across the five distinct challenge categories. The visualizations reveal a high degree of morphological diversity, confirming that Caesar's exploration policy is not static but highly context-dependent, resulting in self-organizing knowledge networks similar to those in \citet{adgr2025}.

The graphs for Constrained Synthesis (Figure~\ref{fig:q1}), Cross-Domain Synthesis (Figure~\ref{fig:q3}), and Meta-Creativity (Figure~\ref{fig:q4}) challenges exhibit a distinct ``starburst'' or high-branching topology. These graphs indicate a breadth-first search strategy where the agent rapidly traverses disjoint semantic clusters to locate novel intersections. For example, in Challenge 3 (``Apply calculus to cooking''), the branching suggests the agent simultaneously explored multiple mathematical sub-domains (e.g., vector fields, accumulation) to find the best structural fit for culinary processes.

Conversely, the graphs for Counterfactual Reasoning (Figure~\ref{fig:q2}) and Open-Ended Synthesis (Figure~\ref{fig:q5}) shift toward long, linear chains with fewer lateral branches and reflect a depth-first exploration strategy instead. For Counterfactual Reasoning, this linear structure likely corresponds to the agent following a specific causal chain (building $C$ upon $B$ upon $A$) to maintain narrative consistency. Similarly, for Open-Ended Synthesis, the depth suggests the agent rapidly selected a promising niche and ``drilled down'' to validate viability, rather than remaining in a shallow brainstorming phase.

Finally, the visualizations illustrate the topological distribution of source nodes cited in the final artifact $A_f$ (colored cyan). The spatial arrangement reveals distinct retrieval patterns, ranging from dense clustering around the center (Figures~\ref{fig:q1},~\ref{fig:q5}) to broad global dispersion (Figures~\ref{fig:q3},~\ref{fig:q4}). Crucially, cited nodes are distributed across varying distances from the root (colored red), with several key sources located near the terminals of extended exploration chains (Figures~\ref{fig:q2},~\ref{fig:q5}). This variance in citation depth confirms that Caesar effectively synthesizes information from the entire trajectory of its deep web exploration, rather than biasing toward early-stage discoveries.

\subsection{Evolution of Knowledge Graphs}
\label{app:temporal_evolution}

\begin{figure}[ht]
    \centering
    \begin{subfigure}[b]{0.32\textwidth}
        \centering
        \includegraphics[width=\linewidth]{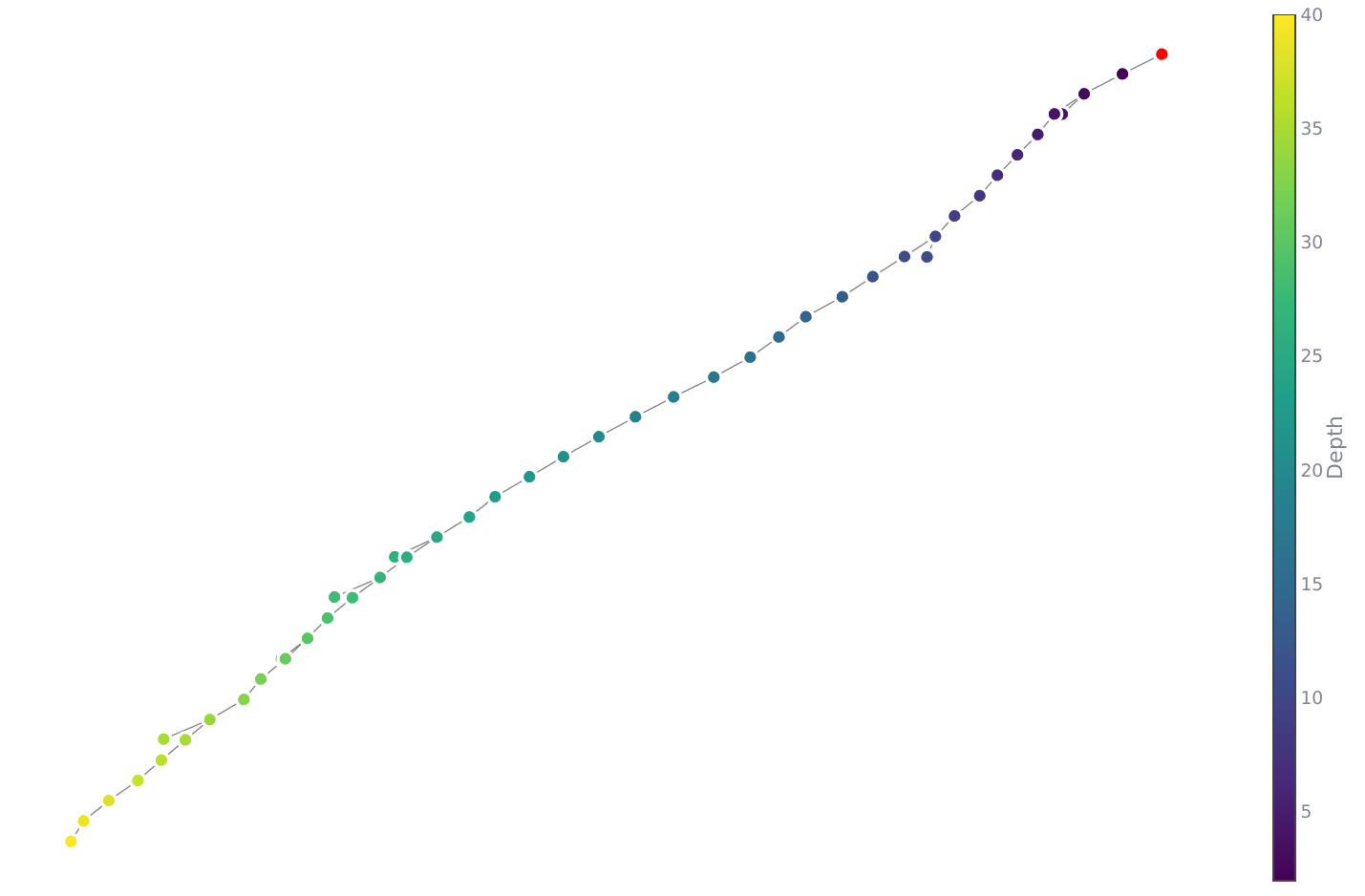}
        \caption{Step 100}
        \label{fig:q5_100}
    \end{subfigure}
    \hfill
    \begin{subfigure}[b]{0.32\textwidth}
        \centering
        \includegraphics[width=\linewidth]{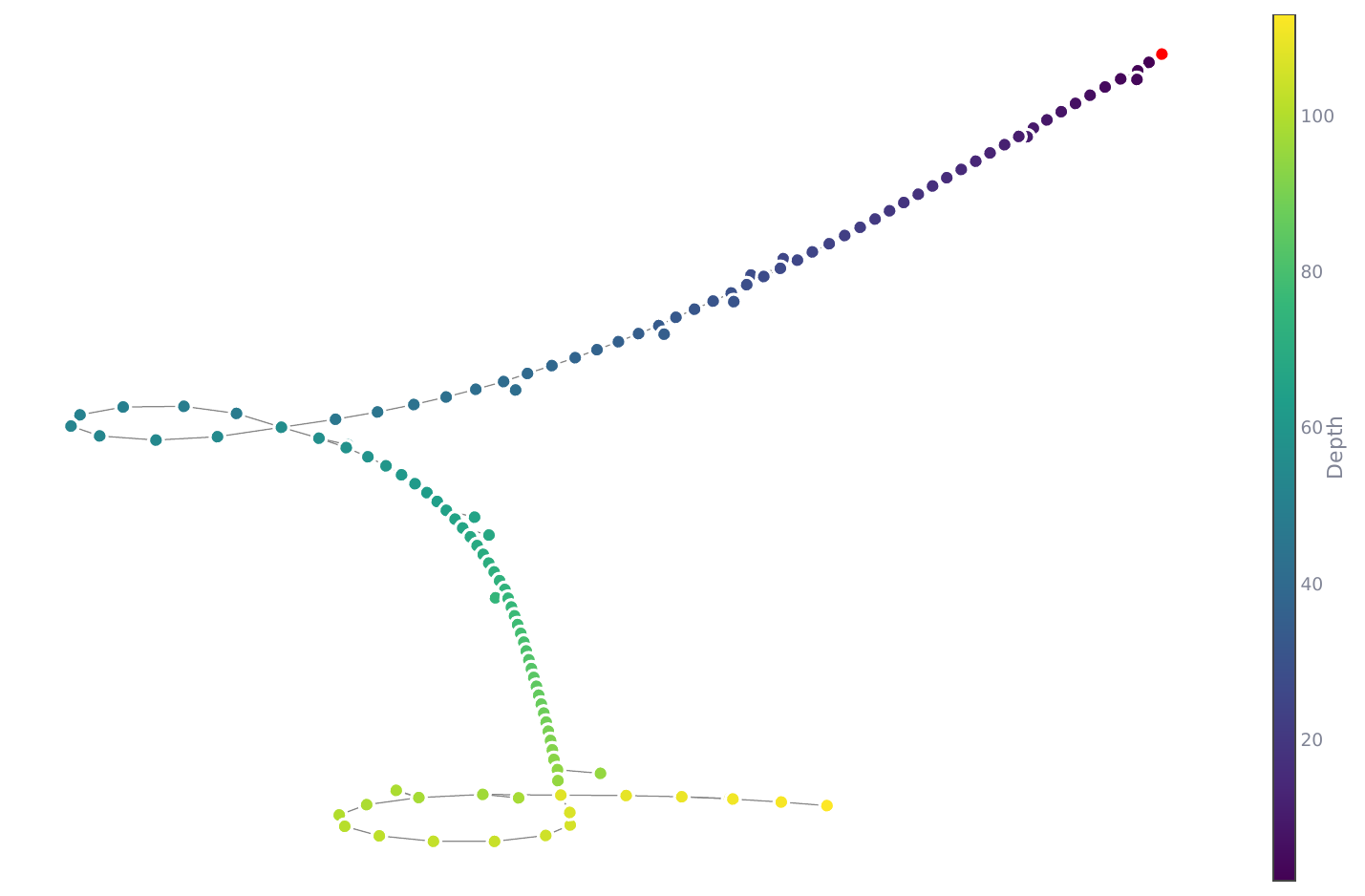}
        \caption{Step 200}
        \label{fig:q5_200}
    \end{subfigure}
    \hfill
    \begin{subfigure}[b]{0.32\textwidth}
        \centering
        \includegraphics[width=\linewidth]{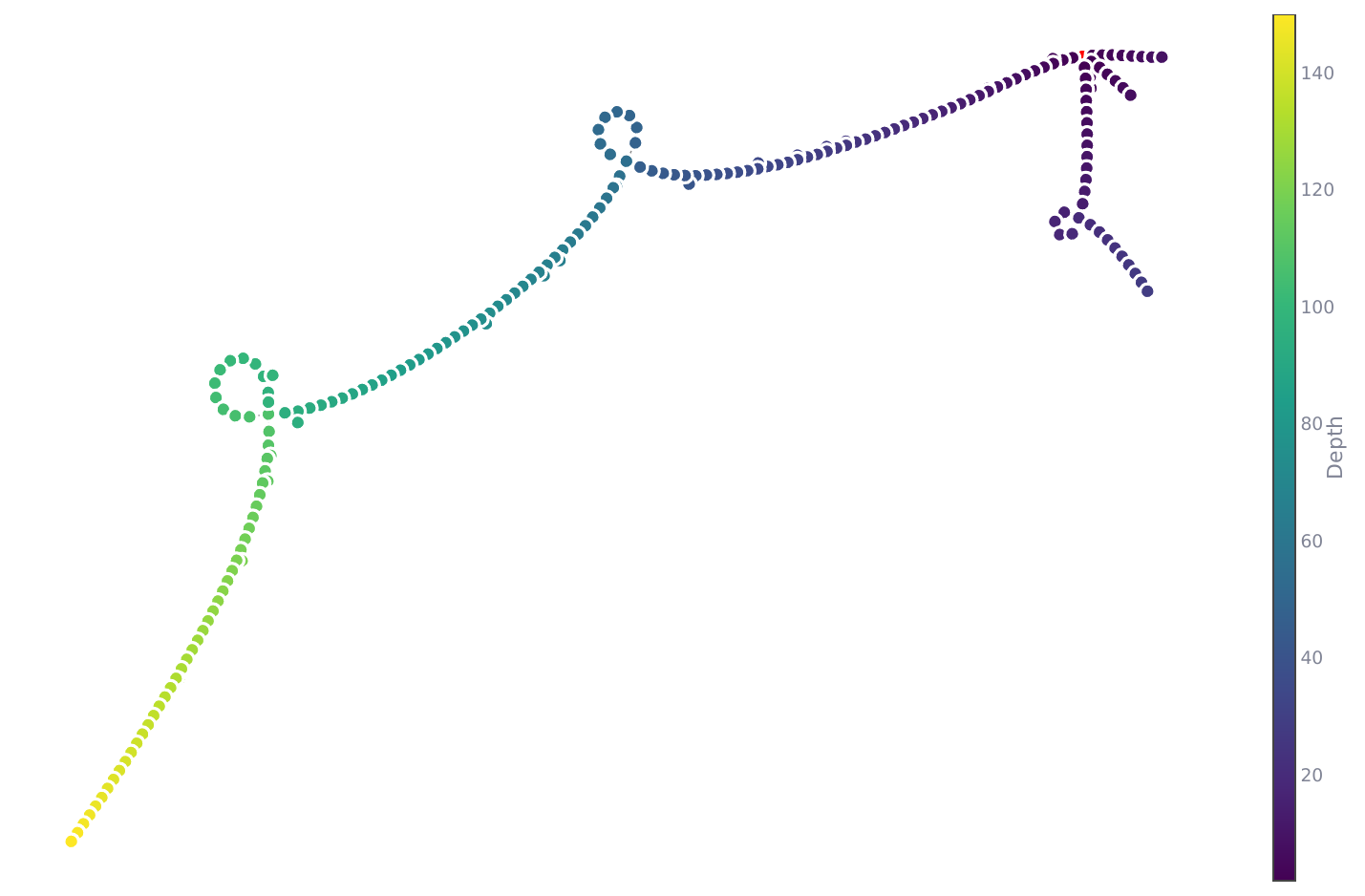}
        \caption{Step 400}
        \label{fig:q5_400}
    \end{subfigure}

    \vspace{1em} 

    \begin{subfigure}[b]{0.32\textwidth}
        \centering
        \includegraphics[width=\linewidth]{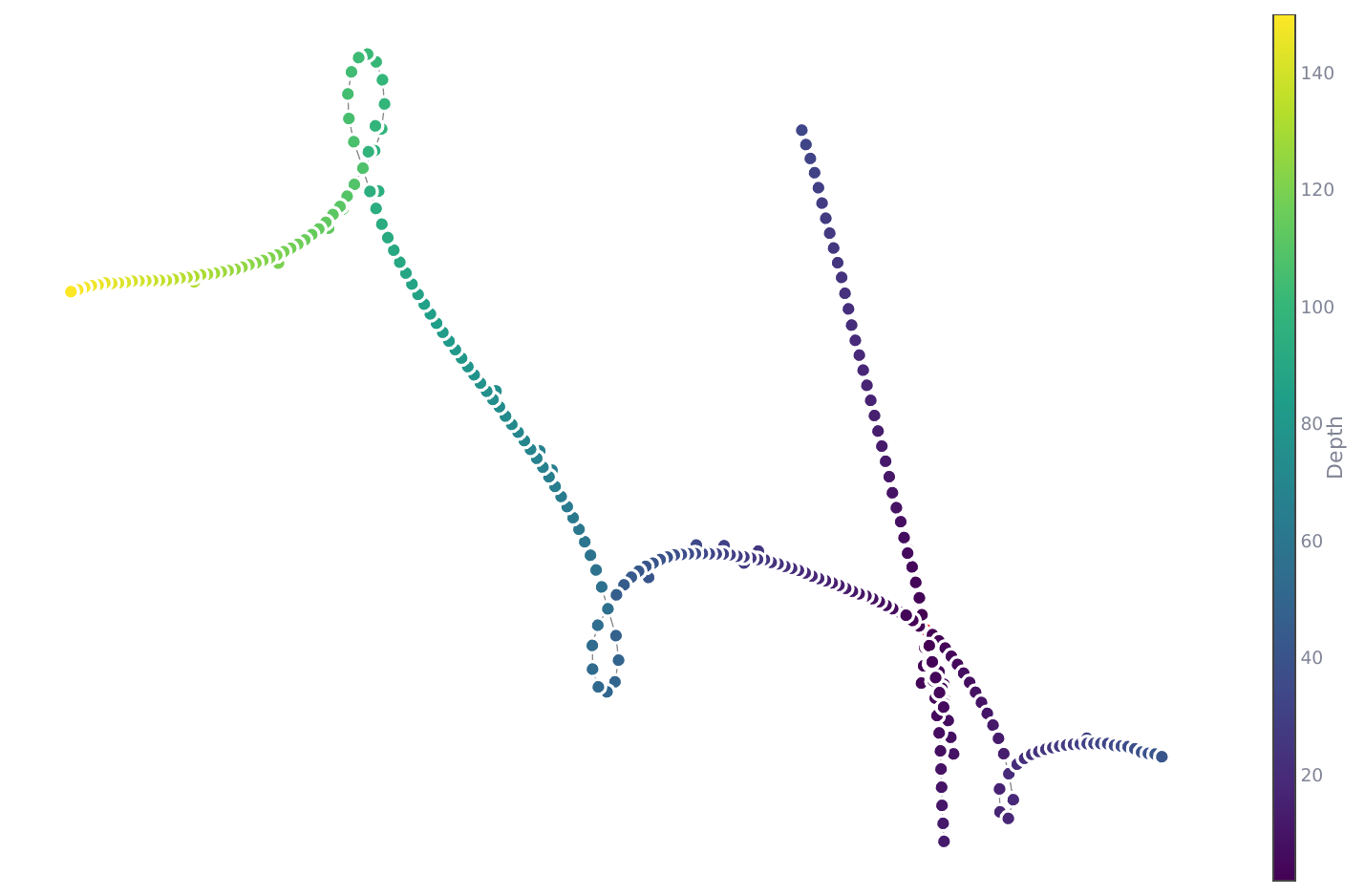}
        \caption{Step 600}
        \label{fig:q5_600}
    \end{subfigure}
    \hfill
    \begin{subfigure}[b]{0.32\textwidth}
        \centering
        \includegraphics[width=\linewidth]{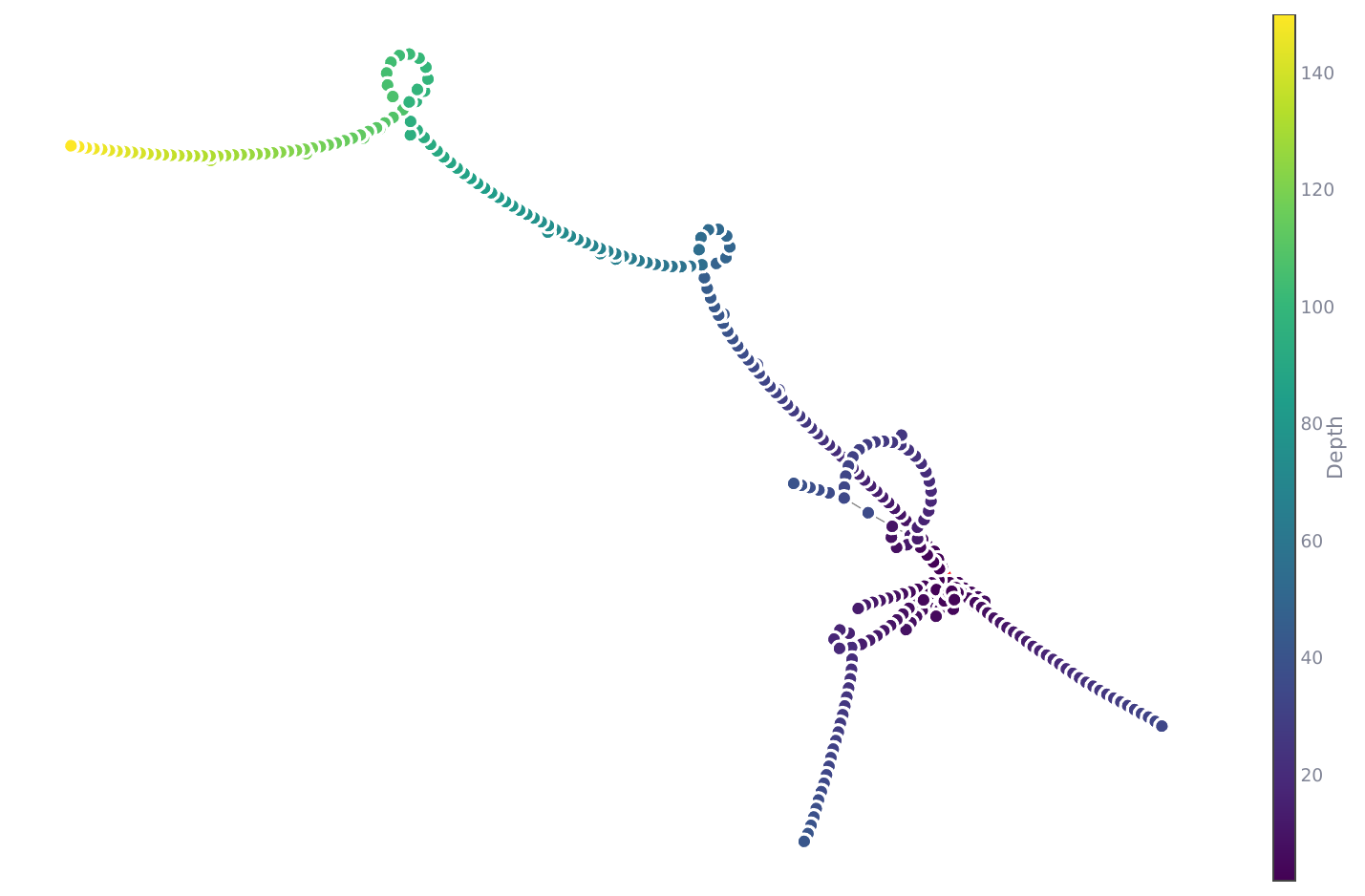}
        \caption{Step 800}
        \label{fig:q5_800}
    \end{subfigure}
    \hfill
    \begin{subfigure}[b]{0.32\textwidth}
        \centering
        \includegraphics[width=\linewidth]{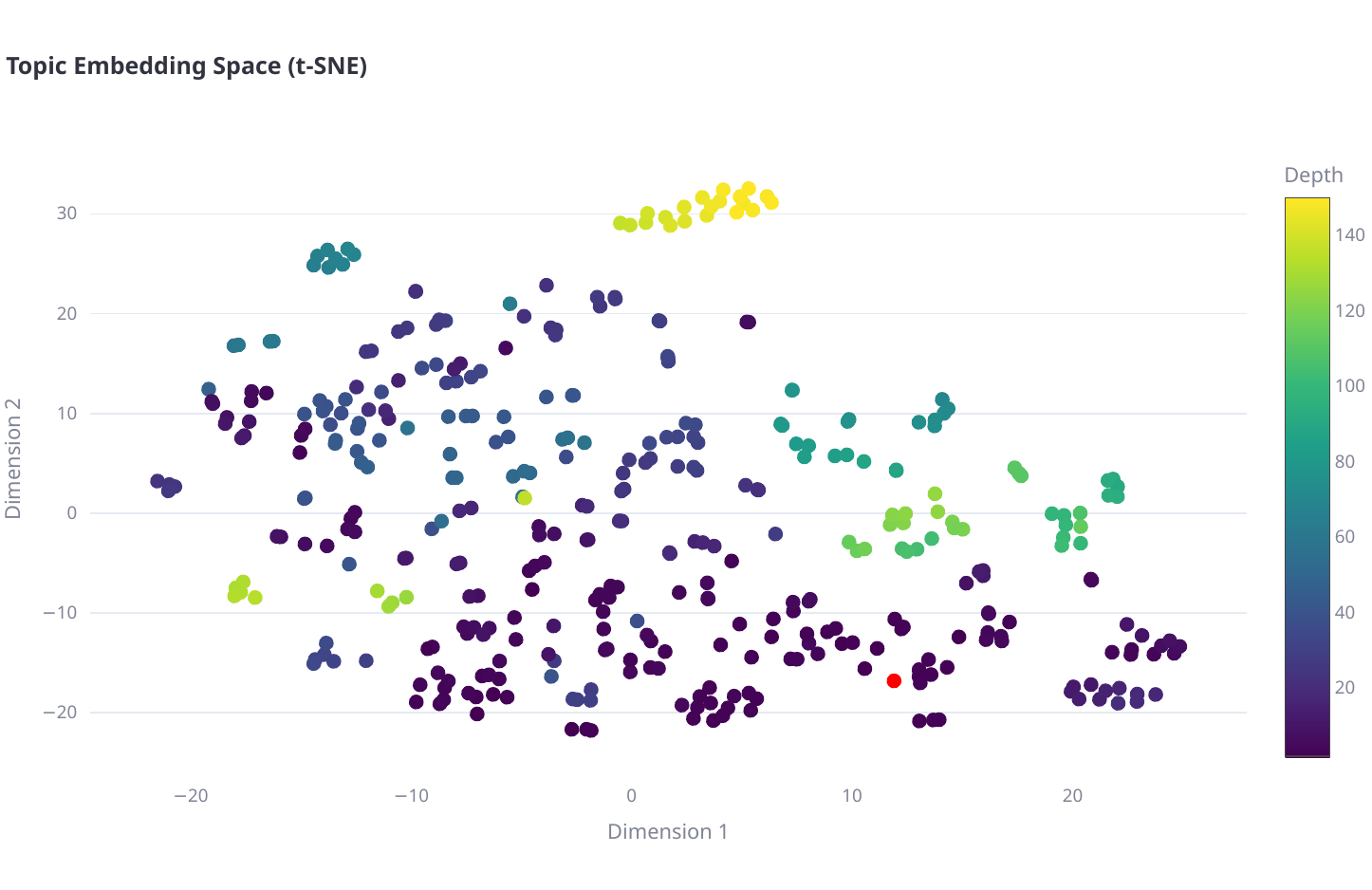}
        \caption{t-SNE Plot}
        \label{fig:tsne}
    \end{subfigure}

    \caption{Evolution of the knowledge graph $G$ for Challenge 5 over 1000 steps. Brighter colors indicate further exploration depth from the root node (red). The figures show a transition from initial depth-first search to breadth-first branching later. The lower right contains a t-SNE \citep{maaten2008visualizing} plot for node text embeddings in $G$ that shows the diversity of insights collected during Caesar's exploration phase.}
    \label{fig:graph_evolution}
\end{figure}

Figure~\ref{fig:graph_evolution} visualizes the step-by-step construction of the knowledge graph $G$ specifically for Challenge 5 (Open-Ended Synthesis). The evolution of the graph over 1000 steps reveals a distinct exploration strategy:

\begin{itemize}[leftmargin=*]
    \item \textbf{Initial Deep Dive (Steps 0-600):} The agent initially pursues a strong depth-first approach, forming a single, increasingly long linear chain (Figures~\ref{fig:q5_100}--\ref{fig:q5_400}).
    \item \textbf{Backtracking and Branching (Steps 600+):} Once this initial path is exhausted, the agent backtracks to around the root node and initiates new, distinct branching chains in different directions (Figures~\ref{fig:q5_600}--\ref{fig:q5_800}).
\end{itemize}

This depth-first, then breadth-next strategy is particularly suited for open-ended ideation tasks. The initial deep dive likely represents the agent thoroughly validating the viability of its first promising lead. By backtracking only after exhausting that specific niche, the agent avoids prematurely converging on a local optimum. The subsequent branching ensures that alternative business concepts are explored, balancing the need for deep validation with the necessity of broad search to find novelty. This strategy mirrors the optimal search dynamic described by \citet{qian2025inforage}, where the agent persists in a specific direction only as long as the information gain remains high, before automatically pivoting to fresh sources once the local insights dry up.

\section{Adversarial Artifact Synthesis Deep Dive}
\label{app:recursive_discovery}
\label{sec:iter_convergence}

This section analyzes the dynamics of Phase 2 (Section~\ref{sec:caesar_synthesis}) and details the operation of the Recursive Insight Discovery mechanism that underpins it. The provided case study illustrates how Caesar constructs a logical dependency chain, iteratively identifying and resolving gaps in its own reasoning to evolve an initial abstract concept into a verified operational model.

\subsection{Mechanism Overview}
\label{app:mechanism}

Underpinning draft synthesis is a recursive insight discovery mechanism that functions as an autonomous interview: the system initializes with the input challenge $Q_0$, retrieves a baseline answer from the knowledge base \emph{KB}, and iteratively poses follow-up questions targeting narrative gaps or contradictions in accumulated answers~\citep{asai2023selfrag}. This depth-first semantic traversal uncovers latent connections often missed by breadth-first synthesis approaches~\citep{souza2025deepresearch}, and the \textsc{MergeDrafts} phase acts as a convergent filter that recovers the \textit{Useful} score lost during adversarial divergence ($7.04 \to 8.28$ in the full answer setting). This two-step process, exploratory divergence followed by strict convergence~\citep{koivisto2023divergent}, is illustrated in the case study below, where Caesar evolves an initial abstract concept into a verified operational model through four iterations of inquiry.

\subsection{Case Study: Open-Ended Synthesis}
\label{app:case_study}

Listing~\ref{lst:recursive_chain} illustrates this mechanism in action for Challenge 5 (``Invent a completely original business''). The reasoning chain demonstrates a progression from high-level conceptualization to granular operational validation:

\begin{itemize}[leftmargin=*]
    \item \textbf{Iteration 1 (Concept):} Introduces the core entity (the ``Resilience Carrier'') as a telecom-like operator that sells continuity of essential services via measurable SLAs.
    \item \textbf{Iteration 2 (Legitimacy \& Governance):} Addresses the trust problem by adding a ``Rights-Holder Data Trust'' (users own the data; Carrier is a fiduciary processor) plus independent SLA verification/auditing.
    \item \textbf{Iteration 3 (Commercial Viability):} Converts the idea into a bounded ``Monsoon Season Readiness'' pilot with a must-cover contract (anti-cherry-picking), a two-part tariff (retainer + triggered payments), and stop-loss/reinsurance-style risk pooling.
    \item \textbf{Iteration 4 (Operations \& Settlement):} Specifies the execution layer (``Service Mesh''): standardized service modules, a tiered supplier registry, real-time dispatch via ``work tokens,'' and escrowed payments released upon proof + verification.
\end{itemize}

This trajectory confirms that Caesar successfully identifies gaps in the current narrative (e.g., ``How do we trust this?'' or ``How do we pay for this?'') and generates adversarial probes to fill them, creating the structural coherence observed in the final artifacts.

\begin{lstlisting}[breaklines=true, basicstyle=\ttfamily\scriptsize, frame=single, caption={Recursive insight discovery chain for Challenge 5 (abridged for readability). The chain evolves from concept to governance to execution, with each question explicitly building upon the previous answer.}, label={lst:recursive_chain}]
[ITERATION 1]
Q: Invent a completely original business idea that doesn't exist yet.
A: The Concept: "Resilience Carrier" (Telecom-style operator for essential services)
   - A for-profit operator selling guaranteed continuity of essential services (healthcare, housing, safety) in volatile regions via Service Level Agreements (SLAs).
   - Core Mechanism: Not an NGO or marketplace, but a network operator accountable for "uptime" (e.g., "95% of households screened within 7 days").
   - User Journey: Families get a "Household Continuity Account" and a Fiduciary Navigator. The operator routes them through a mesh of suppliers to prevent gaps.
   - Business Model: B2G/B2B subscriptions (governments/insurers pay for stability) + Shared-savings (upside from reduced hospitalizations/evictions).

[ITERATION 2]
Q: Given the for-profit nature proposed in Iteration 1, what trust-and-governance mechanism makes this legitimate? Specifically regarding data ownership and SLA verification?
A: Governance Architecture: Rights-Holder Data Trust + SLA Regulator
   1. Data Ownership: The "Household Continuity Account" is owned by the user via a Data Trust. The Carrier is a fiduciary processor with no ownership rights.
   2. Consent under Duress: Features "Duress Modes" releasing only minimum attestations. Violations trigger automatic sanctions.
   3. Verification: An Independent Reliability Regulator (multi-stakeholder board) audits cryptographically signed service receipts. "Silence" is treated as a risk signal.

[ITERATION 3]
Q: With the governance architecture established, what is the smallest pilot product and how is pricing structured to prevent cherry-picking low-risk households?
A: MVP & Financial Structure: "Monsoon Season Readiness" Pilot
   - Pilot Scope: Bounded catchment (e.g., flood corridor) with established providers.
   - Bundle: Navigator access + Mobile Triage + WASH (water/sanitation) repair.
   - SLA Metrics: >90% triage within 2 hours; >85% functional toilet access.
   - Pricing Model:
     1. Geographic "Must-Cover" Contract: Carrier must cover 100% of registered households (no cherry-picking).
     2. Two-Part Tariff: Seasonal Retainer (readiness) + Triggered Usage Payments (surge events).
     3. Stop-loss Pool: Reinsurance-style fund covers costs above catastrophic thresholds.

[ITERATION 4]
Q: To execute the "Monsoon Pilot" model defined above, what is the supply-side operating model? How do you dispatch/pay a heterogeneous network of clinics/vendors in real-time?
A: Operating System: The "Service Mesh"
   1. "Resilience Primitives": Services converted into standardized modules with strict inputs/outputs (e.g., "Acute PTSD Stabilization").
   2. Tiered Registry: From licensed NGOs (Tier A) to community actors (Tier C, sponsored by anchors).
   3. Dispatch Engine: Issues "Work Tokens" based on location, availability, and equity constraints.
   4. Escrow Payment: Two-key release requires Supplier Proof + Independent Verifier confirmation.
   5. Permissioning: Suppliers never own the user record; they write outputs to the ledger via temporary consent tokens.
\end{lstlisting}

\section{Detailed Ablation Results}
\label{app:detailed_ablation}

This section expands upon the ablation results provided in the main paper to better understand how they affect Caesar's drafting processes and ELI5 outputs.

\subsection{Ablation Results for ELI5 Output Constraints}
\label{app:ablation_eli5}

The main paper only reported ablation results for full answers generated by Caesar (Tables~\ref{tab:ablation_graph} and~\ref{tab:ablation_iterations}). Below are the corresponding results for all three output constraints, including the full answer results from the main paper for ease of comparison.

\begin{table}[ht]
\caption{Exploration ablation (\textit{top}) and draft ablation (\textit{bottom}) results across all output constraints.}
\label{tab:ablation_iterations_eli5}
\begin{center}
\begin{footnotesize}
\begin{sc}
\begin{tabular}{lcccc cc}
\toprule
Config & New & Usef. & Surp. & Total & $\delta$ & Effect \\
\midrule
\multicolumn{7}{c}{\textbf{Full Answers}} \\
\midrule
\textbf{1000 Iter} & \textbf{8.22} & \textbf{8.54} & \textbf{8.16} & \textbf{24.92} & -- & -- \\
500 Iter & 7.46 & 7.80 & 7.48 & 22.74 & 0.76 & Large \\
250 Iter & 7.22 & 7.48 & 7.14 & 21.84 & 0.92 & Large \\
\midrule
\multicolumn{7}{c}{\textbf{ELI5 (Unconstrained)}} \\
\midrule
\textbf{1000 Iter} & \textbf{7.56} & \textbf{7.98} & \textbf{7.46} & \textbf{23.00} & -- & -- \\
500 Iter & 7.02 & 7.70 & 6.90 & 21.62 & 0.52 & Large \\
250 Iter & 7.02 & 7.22 & 6.82 & 21.06 & 0.80 & Large \\
\midrule
\multicolumn{7}{c}{\textbf{ELI5 (450 Words)}} \\
\midrule
\textbf{1000 Iter} & \textbf{7.92} & \textbf{8.22} & \textbf{7.82} & \textbf{23.96} & -- & -- \\
500 Iter & 7.24 & 7.78 & 7.18 & 22.20 & 0.48 & Large \\
250 Iter & 6.80 & 7.32 & 6.64 & 20.76 & 1.00 & Large \\
\bottomrule
\end{tabular}

\vspace{1em}

\begin{tabular}{lcccc cc}
\toprule
Variant & New & Usef. & Surp. & Total & $\delta$ & Effect \\
\midrule
\multicolumn{7}{c}{\textbf{Full Answers}} \\
\midrule
\textbf{Final} & \textbf{7.86} & \textbf{8.28} & 7.68 & \textbf{23.82} & -- & -- \\
Draft 3 & 7.90 & 7.04 & \textbf{7.98} & 22.92 & 0.32 & Small \\
Draft 1 & 6.58 & 7.62 & 6.22 & 20.42 & 0.92 & Large \\
\midrule
\multicolumn{7}{c}{\textbf{ELI5 (Unconstrained)}} \\
\midrule
\textbf{Final} & 7.44 & \textbf{7.96} & 7.26 & \textbf{22.66} & -- & -- \\
Draft 3 & \textbf{7.66} & 7.20 & \textbf{7.66} & 22.52 & 0.16 & Small \\
Draft 1 & 6.42 & 7.72 & 6.08 & 20.22 & 0.60 & Large \\
\midrule
\multicolumn{7}{c}{\textbf{ELI5 (450 Words)}} \\
\midrule
Draft 3 & \textbf{7.46} & 7.24 & \textbf{7.68} & \textbf{22.38} & -- & -- \\
\textbf{Final} & 7.10 & \textbf{7.74} & 6.80 & 21.64 & 0.32 & Small \\
Draft 1 & 6.26 & 7.30 & 5.80 & 19.36 & 0.84 & Large \\
\bottomrule
\end{tabular}
\end{sc}
\end{footnotesize}
\end{center}
\end{table}

Notably, under strict length constraints (450 words), Draft 3 ($A_3$) slightly outperforms the final merged artifact ($A_f$), suggesting that compressing diverse perspectives can dilute the narrative sharpness of a focused draft. This trade-off does not appear in the unconstrained settings.

\subsection{Full Knowledge Graph Ablation Results}
\label{app:graph_ablation_full}

Table~\ref{tab:graph_ablation_full} presents the knowledge graph ablation results broken down by draft iteration, extending the summary in Table~\ref{tab:ablation_graph} of the main paper. These results demonstrate how the graph's contribution manifests across the adversarial refinement process.

\begin{table}[ht]
\caption{Knowledge graph ablation by draft iteration (Full Answers). Cliff's Delta is computed against the corresponding Control draft. The No-KG ablation (replacing graph-guided exploration with flat top-$T$ search) degrades creativity most strongly in Draft 1 and the Final merge.}
\label{tab:graph_ablation_full}
\begin{center}
\begin{small}
\begin{sc}
\begin{tabular}{lcccc}
\toprule
Draft & Control & No KG & $\delta$ & Effect \\
\midrule
Draft 1 & 24.45 & 23.23 & 0.56 & Large \\
Draft 2 & 24.37 & 23.84 & 0.28 & Small \\
Draft 3 & 24.45 & 23.81 & 0.28 & Small \\
Final Merge & 25.11 & 23.76 & 0.52 & Large \\
\bottomrule
\end{tabular}
\end{sc}
\end{small}
\end{center}
\end{table}

The draft-level analysis reveals that the knowledge graph has the strongest effect on Draft 1 ($\delta = 0.56$) and the Final merged artifact ($\delta = 0.52$). The intermediate drafts show smaller effects ($\delta = 0.28$), likely because the adversarial refinement loop partially compensates for the weaker initial knowledge base through its iterative questioning mechanism. However, the Final merge, which synthesizes across all drafts, cannot fully recover the lost diversity, resulting in a large overall effect.

Table~\ref{tab:multihop_ablation} presents the single-hop ablation results, showing the effect of decreasing $\mathcal{N}_c$ to $k{=}1$ (5 nodes max) from $k{=}3$ (30 nodes max).

\begin{table}[ht]
\caption{Single-hop neighborhood ablation by draft iteration (Full Answers). Decreasing $\mathcal{N}_c$ to $k{=}1$ causes consistent performance degradation over the standard $k{=}3$ configuration.}
\label{tab:multihop_ablation}
\begin{center}
\begin{small}
\begin{sc}
\begin{tabular}{lcccc}
\toprule
Draft & Control & Single-Hop & $\delta$ & Effect \\
\midrule
Draft 1 & 24.75 & 23.49 & 0.68 & Large \\
Draft 2 & 24.80 & 23.71 & 0.32 & Small \\
Draft 3 & 24.88 & 23.72 & 0.44 & Medium \\
Final Merge & 25.45 & 23.69 & 0.52 & Large \\
\bottomrule
\end{tabular}
\end{sc}
\end{small}
\end{center}
\end{table}

The single-hop ablation shows the largest effect in Draft 1 ($\delta = 0.68$), suggesting that richer graph context is most beneficial during initial synthesis when the agent must establish its creative framework. The Final merge also shows a large effect ($\delta = 0.52$), confirming that the broader neighborhood context provides lasting benefits.

To further understand why graph structure matters, creativity scores were compared between the top 10 hub nodes (high connectivity) and bottom 10 leaf nodes (low connectivity) for the $G$ that was explored for each challenge. Overall, hub node insights scored significantly higher than leaf insights (24.43 vs. 21.88, $\delta = 1.00$). This indicates that hub nodes, which aggregate information from diverse branches, are particularly valuable for creative synthesis.

\section{Compute-Controlled Comparison}
\label{app:cost_controlled}

To address the confound of uncontrolled compute budgets across baselines, a cost-controlled experiment was conducted. Caesar was limited to $T{=}250$ exploration steps using GPT-5-mini for Phase 1, matching the estimated \$5 per-challenge cost of Gemini Deep Research. The evaluation follows the same protocol as Table~\ref{tab:main_results} under the unconstrained full-answer setting. As shown in Table~\ref{tab:cost_controlled}, Caesar is able to outperform all of the deep research baselines with large values for $\delta$.

\begin{table}[ht]
\caption{Cost-controlled comparison. Caesar at $T{=}250$ with GPT-5-mini matches the estimated compute budget of Gemini Deep Research while still outperforming all baselines with large effect sizes.}
\label{tab:cost_controlled}
\begin{center}
\begin{small}
\begin{sc}
\begin{tabular}{lccc}
\toprule
Deep Agent & Avg Score & $\delta$ & Effect \\
\midrule
\textbf{Caesar} ($T{=}250$, GPT-5-\textnormal{mini}) & \textbf{26.16} & -- & -- \\
Gemini 3 (Deep) & 24.37 & 0.76 & Large \\
Sonnet 4.5 (Deep) & 21.00 & 1.00 & Large \\
GPT-5.2 (Deep) & 16.16 & 1.00 & Large \\
\bottomrule
\end{tabular}
\end{sc}
\end{small}
\end{center}
\end{table}

Even under this strict budget constraint, Caesar significantly outperforms all deep research baselines. This result demonstrates that Caesar's architectural advantage (graph-guided exploration + adversarial synthesis) is not merely a function of increased compute expenditure, but reflects genuine algorithmic improvements in how the exploration budget is utilized.

\section{Human Evaluation Study}
\label{app:human_eval}

To validate the LLM-as-a-Judge framework, a human evaluation study was conducted comparing Caesar against the strongest baseline (Gemini 3 Deep Research, which was rated second best by the LLM judges).

\subsection{Study Design}
23 human raters were recruited to evaluate pairwise comparisons between Caesar and the baseline. The full answers of both agents were passed through the same LLM-based normalization step (Listing~\ref{lst:prompt_human_eval}) that rewrites each answer into a fixed two-paragraph schema: a 2--3 sentence plain-language summary of the core idea (prefixed \texttt{SUMMARY:}) and a 3--4 sentence argument for why it is creative (prefixed \texttt{WHY IT'S CREATIVE:}). This format was explicitly designed to allow raters to quickly judge the underlying idea while not getting distracted by surface verbosity or structure. Human judges were then asked to select which of two anonymized normalized answers was more creative according to the NUS rubric used by the LLM judges. Each of the five challenges produced multiple comparison pairs, yielding 112 total A/B matchups.

\begin{lstlisting}[breaklines=true, basicstyle=\ttfamily\small, frame=single, caption={Human Evaluation Normalization Prompt}, label={lst:prompt_human_eval}]
ARTIFACT: {artifact_text}

YOUR TASK:
Write exactly two short paragraphs about the artifact above.

PARAGRAPH 1 - CORE IDEA SUMMARY (2-3 sentences):
- Identify and summarize the most important creative key idea in the artifact
- Make sure the summary is clear and understandable to a non-expert
- The reader should immediately understand what the artifact is about
- Do not overuse jargon or too many technical terms

PARAGRAPH 2 - CREATIVITY ARGUMENT (3-4 sentences):
- Make a clear, convincing argument for why the key idea is creative
- Explain what makes it novel, surprising, or useful
- Ground your argument in specific aspects of the idea, not generic praise
- Write to persuade a skeptical reader

IMPORTANT: Do NOT mention or reference the artifact; write as if describing the idea itself.
IMPORTANT: Start paragraph 1 with "SUMMARY: " and paragraph 2 with "WHY IT'S CREATIVE: "
\end{lstlisting}

\subsection{Results}

\begin{table}[ht]
\caption{Human evaluation results: pairwise preference counts for Caesar vs. Gemini 3 Deep Research across all five challenges (C1: Constrained Synthesis, C2: Counterfactual Reasoning, C3: Cross-Domain Synthesis, C4: Meta-Creativity, C5: Open-Ended Synthesis).}
\label{tab:human_eval}
\begin{center}
\begin{small}
\begin{sc}
\begin{tabular}{lcccccc}
\toprule
Model & C1 & C2 & C3 & C4 & C5 & Total \\
\midrule
Caesar & 5 & 18 & 20 & 13 & 7 & \textbf{63} \\
Gemini 3 & 18 & 5 & 2 & 9 & 15 & 49 \\
\bottomrule
\end{tabular}
\end{sc}
\end{small}
\end{center}
\end{table}

Human raters preferred Caesar overall (63 vs. 49, 56.25\% preference rate, odds ratio 1.29). The pattern across challenges mirrors the LLM judge results: Caesar is strongly preferred for tasks requiring cross-domain synthesis (C3: 20 vs. 2) and counterfactual reasoning (C2: 18 vs. 5), while the preference is reversed for constrained synthesis (C1: 5 vs. 18) and open-ended synthesis (C5: 7 vs. 15). This alignment between human and LLM judge evaluations supports the reliability of the automated scoring framework.

\section{Caesar Implementation Details}
\label{app:implementation}

This section outlines the technical specifications of the Caesar architecture, detailing the software stack used for graph management and web perception. It also provides the set of hyperparameters configured for the exploration and synthesis phases to ensure reproducibility.

\subsection{System Architecture}
The Caesar agent framework was implemented using the Python programming language. The core components that use external libraries are:

\begin{itemize}[leftmargin=*]
    \item \textbf{Graph Management:} The navigational graph $G$ is managed using \texttt{NetworkX} \citep{hagberg2008networkx}. Each node represents a visited URL, containing attributes for the raw text content, the timestamp of access, and the depth relative to the root node. The graph is directed, with edges representing navigational transitions (e.g., clicking a link).
    \item \textbf{Vector Store:} The store utilizes \texttt{ChromaDB} \citep{chroma2024} for the knowledge base \emph{KB}. Text chunks are embedded using \texttt{text-embedding-3-large} (via OpenAI) with a chunk size of 400 tokens and an overlap of 80 tokens. The system supports metadata filtering based on iteration and depth.
    \item \textbf{Web Perception:} To mitigate anti-bot measures, a custom \texttt{curl\_cffi} wrapper (configured to impersonate Chrome) was used to fetch HTML content. This wrapper handles TLS fingerprinting, automatic decompression, and header management to mimic legitimate browser traffic. Content extraction and filtering of irrelevant text (e.g., SEO links) is handled by \texttt{Trafilatura} \citep{barbaresi2021trafilatura} for HTML and \texttt{PyPDF2} \citep{pypdf2} for PDF documents.
    \item \textbf{LLM Backend:} The experiments were conducted using GPT-5.2 \citep{openai2025gpt52} as the primary driver for both exploration and synthesis. The system uses low reasoning effort during the exploration phase to increase throughput and high reasoning effort during the artifact synthesis phase to ensure quality of answers.
    \item \textbf{Memory Layer:} An exploration memory store was implemented using the \texttt{Mem0} library \citep{singh2024mem0}, which integrates a \texttt{ChromaDB} vector store with a \texttt{Neo4j} \citep{neo4j_2024} graph database for managing long-term agent memory and detecting navigational loops.
\end{itemize}

\subsection{Hyperparameters}
\label{app:hyperparameters}

Table~\ref{tab:hyperparams} details the specific hyperparameters used for the main experiments reported in Section~\ref{sec:experiments}.

\begin{table}[H]
\caption{Hyperparameter settings for both web exploration (Section~\ref{sec:caesar_explore}) and artifact synthesis (Section~\ref{sec:caesar_synthesis}) phases of Caesar.}
\label{tab:hyperparams}
\begin{center}
\begin{scriptsize}
\begin{sc}
\begin{tabular}{llc}
\toprule
Parameter & Description & Value \\
\midrule
\multicolumn{3}{c}{\textbf{Phase 1: Deep Web Exploration}} \\
\midrule
$T$ & Exploration Budget (Steps) & 1000 \\
$P_m$ & Max Page Content (Char) & 100k \\
$L_m$ & Max Candidate Links per Page & 2000 \\
$R_m$ & Max Allowed Page Revisits & 20 \\
$S_m$ & Max Web Search Actions & 30 \\
$D_m$ & Max Graph Exploration Depth & 10000 \\
$|\mathcal{N}_c|$ & Max Graph Neighbor Context Size ($k{=}3$ hops) & 30 \\
$\tau_e$ & LLM Temperature for Think/Act & 0.9 \\
$R_e$ & LLM Reasoning Effort for Think/Act & Low \\
\midrule
\multicolumn{3}{c}{\textbf{Phase 2: Adversarial Artifact Synthesis}} \\
\midrule
$\hat{T}$ & Recursive Insight Budget (Iter) & 30 \\
$N$ & Adversarial Refinement Rounds & 3 \\
$H_c$ & Max QA Context History & 50 \\
$C_m$ & Max Citations per Claim & 5 \\
$\tau_s$ & LLM Temperature for Draft/Merge & 0.1 \\
$R_s$ & LLM Reasoning Effort for Draft/Merge & High \\
\midrule
\multicolumn{3}{c}{\textbf{Global Settings}} \\
\midrule
$O_m$ & Max LLM Output (Tokens) & 50k \\
$R_k$ & Top-$k$ KB Retrieval & 50 \\
$R_n$ & Top-$n$ KB Reranking & 10 \\
\bottomrule
\end{tabular}
\end{sc}
\end{scriptsize}
\end{center}
\end{table}

\section{Prompts for Caesar}
\label{app:core_prompts}

This section shows the LLM instructions from both the exploration and synthesis phases of Caesar. These prompts (abridged for readability) demonstrate how Caesar moves beyond standard RAG retrieval by carefully incorporating information from the knowledge graph and past drafts as context.

\subsection{Phase 1 Prompts (Deep Web Exploration)}

\textbf{Graph-Augmented Insight Generation (Think).} The LLM evaluates the relationship between the current page content and insights from local knowledge graph nodes. Rather than summarizing the page in isolation, Caesar identifies narrative gaps and unexpected connections using the insights as context.

\begin{lstlisting}[breaklines=true, basicstyle=\ttfamily\small, frame=single, caption={Graph-Augmented Insight Generation Prompt}, label={lst:prompt_think}]
PAGE CONTENT: {page_content}

INITIAL QUERY: {initial_query}

PAST INSIGHTS: {past_insights}

NEIGHBOR INSIGHTS: {neighbor_insights}

YOUR TASK:
Analyze this content and extract key insights focusing on:
- Novel patterns or unexpected connections
- Assumptions being made and alternative perspectives
- Interesting questions raised by the content
- How to answer the query
- How this builds upon or challenges past/neighbor insights

Depending on the complexity of the content, provide anywhere from 1 to 6 concise but substantive insights, but do not exceed ~600 words in total length:
\end{lstlisting}

\textbf{Knowledge-Guided Exploration (Act).} To prevent navigational amnesia, a high-level meta-strategy (\textsc{Explore}, \textsc{Backtrack}, or \textsc{WebSearch}) is determined based on the current state of exploration. Afterwards, the next link is selected based on that strategy.

\begin{lstlisting}[breaklines=true, basicstyle=\ttfamily\small, frame=single, caption={Knowledge-Guided Exploration Prompt}, label={lst:prompt_meta_strategy}]
CURRENT EXPLORATION CONTEXT:
- Current step: {current_step}/{max_steps}
- Current depth: {current_depth}/{max_depth}
- Web pages visited: {len(visited_urls)}
- Current URL: {current_url}

CURRENT EXPLORATION INSIGHTS:
{kb_context if kb_context else "No exploration insights available."}

HISTORICAL NAVIGATION PATTERNS:
{memory_context if memory_context else "No exploration history available."}

Analyze whether the agent should:
1. **EXPLORE** new un-visited pages to discover novel information or knowledge
2. **BACKTRACK** to the immediate previously visited page to try alternative paths
3. **WEB_SEARCH** relevant topics to address current exploration insights

Consider:
- Knowledge gaps vs areas of saturation
- Depth of current exploration branch
- Success patterns from previous decisions
- Risk/reward of new exploration vs consolidation
\end{lstlisting}

\subsection{Phase 2 Prompts (Adversarial Artifact Synthesis)}

\textbf{Recursive Insight Discovery.} During the initial insight discovery stage, the agent generates the next logical question in the chain of thought using the previous question/answer insights as context.

\begin{lstlisting}[breaklines=true, basicstyle=\ttfamily\small, frame=single, caption={Recursive Insight Discovery Prompt}, label={lst:prompt_recursive}]
PREVIOUS INSIGHTS: {list_of_qa_insights}

YOUR TASK:
Based on the insights gathered so far, what is the next most important question to ask
to deepen understanding and reveal emergent patterns? The question should:
- Build on previous insights rather than repeat them
- Seek connections between different themes
- Identify gaps or contradictions to explore
- Move toward synthesis and creation rather than enumeration
\end{lstlisting}

\textbf{Recurrent Draft Generation.} In the multi-draft loop, this prompt generates the artifact. Crucially, if a previous draft exists, the prompt injects it into the context and explicitly instructs the LLM to critique and improve upon the prior work.

\begin{lstlisting}[breaklines=true, basicstyle=\ttfamily\small, frame=single, caption={Recurrent Draft Generation Prompt}, label={lst:prompt_draft_generation}]
KEY INSIGHTS: {list_of_qa_insights}

PREVIOUS ARTIFACT: {artifact_text}

YOUR TASK:
Drawing heavily upon the patterns that emerged from the key insights, and building upon the previous artifact, create a novel, exciting, and thought provoking artifact that creatively answers this query: {starting_query}
- Emergent patterns not visible in individual sources
- Novel discoveries, connections, or applications
- Surprising new directions or perspectives
- Interesting tensions, contradictions, or open questions

IMPORTANT: do NOT mention or reference the previous artifact, the new artifact should make sense by itself as a standalone text.
IMPORTANT: Avoid excessive jargon, ensure artifact text is well-organized (logical, clear, focused), and convincing to a skeptical reader
\end{lstlisting}

\textbf{Adversarial Challenge Refinement.} Between drafts, the LLM is prompted to analyze the previous artifact text for weaknesses and formulate a new, refined challenge that targets these weaknesses for the next round of synthesis.

\begin{lstlisting}[breaklines=true, basicstyle=\ttfamily\small, frame=single, caption={Adversarial Challenge Refinement Prompt}, label={lst:prompt_refine}]
PREVIOUS QUERY: {previous_query}

PREVIOUS ARTIFACT: {artifact_text}

YOUR TASK:
Based on the previous query and artifact above, identify the most promising direction for deeper exploration. What NEW question or angle would:
- Build on the insights already discovered
- Explore gaps, contradictions, or unexplored connections
- Lead to novel perspectives or applications
- Go deeper rather than broader

The refined query should be concise (1-2 sentences), straightforward, clear, and understandable.
\end{lstlisting}

\textbf{Generative Draft Merging.} Finally, the LLM merges all rounds of artifact drafts into a single coherent narrative, identifying tensions and unifying perspectives.

\begin{lstlisting}[breaklines=true, basicstyle=\ttfamily\small, frame=single, caption={Generative Draft Merging Prompt}, label={lst:prompt_merge}]
ARTIFACT DRAFTS: {list_of_artifact_drafts}

YOUR TASK:
Create a comprehensive merged artifact that:
- Combines the draft artifacts into a single cohesive and complete artifact
- Selectively integrates the most interesting, relevant insights across all draft artifacts
- Discovers emergent patterns not visible in individual artifacts
- Further develops the core strengths while addressing the weaknesses of the draft artifacts
\end{lstlisting}

\textbf{Post-Processing and Summarization.} The post-processing step distills the final merged artifact $A_f$ into ELI5 language using the prompt shown in Listing~\ref{lst:eli5_answer_prompt} (Section~\ref{app:experiment_setup}); the same prompt is applied uniformly to all agents during evaluation.

\section{Computational Cost Analysis}
\label{app:cost}

While Caesar achieves higher creativity scores, its heavy usage of LLMs for web page processing and vector store retrieval incurs a noticeable computation cost. The total cost is split between the extensive data gathering in Phase 1 and the recursive draft generation in Phase 2. The computational cost suggests that Caesar is best utilized as a tool for high-value, asynchronous research rather than real-time interaction, similar to the deep research agents described in \citet{li2025webthinker}.

\subsection{Phase 1: Deep Web Exploration Costs}

The exploration phase constitutes the primary computational expense, averaging approximately \$40 to \$60 per 1000-step experiment. Given the GPT-5.2 pricing of \$1.75 per 1M input tokens and an average context load of $\sim$30k tokens per step, this cost reflects the cumulative impact of continuous context usage and output generation during web exploration. The token budget is predominantly consumed by two mechanisms:

\begin{itemize}[leftmargin=*]
    \item \textbf{Ingestion of Text-Heavy Documents:} Caesar processes most or all of the text content of a page to extract deep insights. In most cases, Caesar processes standard web articles with only $\sim$5k tokens. However, when exploring scientific and humanitarian domains, the agent can encounter dense PDF reports and data portals. These text-heavy pages can reach the truncated limit of $P_m$ (100k characters), consuming 20k--25k tokens for the page content alone.

    \item \textbf{Processing and Selecting Links:} The other major consumer of the token budget is the link selection in the ``Act'' stage. To make informed navigational decisions, the agent processes hundreds of links per page on average (ranging from $\sim$500 on wiki-style pages to 1000+ on dense index pages). Assuming each link consumes $\sim$25 tokens of context, this results in an overall cost of $\sim$10 to 20k tokens per navigation step, regardless of the remaining page content.
\end{itemize}

\subsection{Phase 2: Adversarial Artifact Synthesis Costs}
In contrast, the synthesis phase is comparatively inexpensive, averaging less than \$3 per round.

\begin{itemize}[leftmargin=*]
    \item \textbf{Efficient Insight Retrieval:} Unlike the exploration phase, which ingests raw uncompressed web content, the synthesis phase operates on a curated knowledge base (\emph{KB}). By limiting the number of insights retrieved and performing reranking on them to generate concise answers, the context window is significantly reduced, minimizing input token costs.
    \item \textbf{High Reasoning Overhead:} The cost driver in this phase shifts to output tokens. The synthesizer utilizes GPT-5.2 with high reasoning effort during the adversarial refinement rounds ($N=3$). While the input volume is low, the generation of extensive chain-of-thought reasoning \citep{long2023tot} to integrate disparate insights increases the cost per inference step relative to standard generation.
\end{itemize}

\section{Broader Impacts and Responsible Use}
\label{app:broader_impacts}

The research presented in this paper advances the capabilities of autonomous agents in synthesizing complex information from the open web. However, the architecture introduced in Caesar, specifically the decoupling of graph-based exploration from artifact synthesis, necessitates careful consideration of societal risks.

First, the deployment of recursive web agents raises concerns regarding infrastructure strain and data privacy. While Caesar utilizes browser fingerprinting to ensure robust compatibility with modern web rendering standards during benchmarking, responsible real-world deployment requires strict adherence to \texttt{robots.txt} exclusion protocols and domain-level rate-limiting to preserve site integrity.

Second, the system's objective to maximize creativity introduces a dual-use risk regarding the generation of persuasive misinformation. While the proposed citation mechanism (Section~\ref{sec:caesar_synthesis}) is designed to enforce provenance, the potential for agents to construct compelling but factually tenuous narratives remains an issue. Consequently, this framework is intended as a human-in-the-loop research support tool rather than as a fully autonomous content generator, underscoring the need for continued development of verifiable constraints for creative generation.

In accordance with relevant policies on the use of generative AI, this work utilized LLMs to assist in the generation of Python code for the experimental framework and to improve the clarity and grammatical flow of the manuscript. However, all scientific claims, algorithm/experimental designs, verification of results, and creation of the written content were done by the human authors.

\end{document}